\documentclass[10pt]{ieeetj}
\usepackage{cite}
\usepackage{amsmath,amssymb,amsfonts,amsthm,comment,multirow}
\usepackage{algorithmic}
\usepackage{graphicx,color,bm,url,subfigure}
\usepackage{textcomp}
\usepackage{xcolor}
\usepackage{hyperref}
\hypersetup{hidelinks=true}
\usepackage{algorithm,algorithmic}
\def\BibTeX{{\rm B\kern-.05em{\sc i\kern-.025em b}\kern-.08em
    T\kern-.1667em\lower.7ex\hbox{E}\kern-.125emX}}
\AtBeginDocument{\definecolor{tmlcncolor}{cmyk}{0.93,0.59,0.15,0.02}\definecolor{NavyBlue}{RGB}{0,86,125}}

\makeatletter
\@namedef{revisionnewref@rampa2015physical}{1}
\@namedef{revisionnewref@rampa2017em}{1}
\@namedef{revisionnewref@rampa2022arrays}{1}
\@namedef{revisionnewref@rampa2022multiple}{1}
\@namedef{revisionnewref@wax1985detection}{1}
\@namedef{revisionnewref@ITURP526}{1}
\newcommand{\revisionbibcolor}[1]{%
  \@ifundefined{revisionnewref@#1}{\color{black}}{\color{black}}}
\AtBeginDocument{%
  \let\revisionoriginalbibitem\bibitem
  \renewcommand{\bibitem}{\@ifnextchar[{\revisionbibitemopt}{\revisionbibitemplain}}%
}
\def\revisionbibitemplain#1{\revisionoriginalbibitem{#1}\revisionbibcolor{#1}}
\def\revisionbibitemopt[#1]#2{\revisionoriginalbibitem[#1]{#2}\revisionbibcolor{#2}}
\makeatother

\def\OJlogo{\vspace{-4pt}$<$Society logo(s) and publication title will appear here.$>$}
\def\seclogo{\vspace{10pt}$<$Society logo(s) and publication title will appear here.$>$}

\def\authorrefmark#1{\ensuremath{^{\textbf{#1}}}}

\newtheorem{theorem}{Theorem}

\begin{document}
\receiveddate{XX Month, XXXX}
\reviseddate{XX Month, XXXX}
\accepteddate{XX Month, XXXX}
\publisheddate{XX Month, XXXX}
%\currentdate{XX Month, XXXX}
\doiinfo{XXXX.2022.1234567}

\markboth{}{Author {et al.}}

\title{\textcolor{black}{Joint Source Localization and Environmental Sensing using Parametric Diffraction Array Manifolds}}

\author{Jiaqi Xu\authorrefmark{1}, Bj\"orn Ottersten\authorrefmark{2}, Fellow, IEEE, and A. Lee Swindlehurst\authorrefmark{1}, Life Fellow, IEEE}
\affil{Department of Electrical Engineering \& Computer Science, University of California Irvine, Irvine, CA 92617 USA}
\affil{Interdisciplinary Centre for Security, Reliability and Trust, University of Luxembourg, L-1855 Luxembourg}
%\affil{Electrical Engineering Department, University of Colorado, Boulder, CO 80309 USA}
\corresp{Corresponding author: A. Lee Swindlehurst (email: swindle@uci.edu).}
%\authornote{This paragraph of the first footnote will contain support information, including sponsor and financial support acknowledgment. For example, ``This work was supported in part by the U.S. Department of Commerce under Grant 123456.''}

%\maketitle

\begin{abstract}
\textcolor{black}{We formulate the problem of environmental-object sensing as a joint source-localization and parameter-estimation problem using a diffraction-aware array manifold. The goal is a computationally manageable bridge between conventional array manifolds and full-wave electromagnetic simulation. Unlike prior parametric diffraction studies focused primarily on received-signal-strength measurements across radio links, the proposed formulation exploits the complex spatial array response to jointly infer source and environmental parameters. We employ a scalar diffraction model with obliquity correction that accommodates both plane-wave and spherical-wave illumination. Its thin-screen form admits Fresnel and Fraunhofer simplifications, but we also consider finite-thickness object models that include the approximate effect of dielectric transmission and bulk propagation. We consider various object shape parameterizations, stochastic maximum-likelihood (ML) estimation, Cram\'er--Rao bounds (CRBs), and model-order selection. Numerical experiments address joint source and object parameter estimation, and the resulting RMSEs are shown to approach the associated stochastic CRBs. Comparisons with simplified blockage representations illustrate their potential mismatch bias.}
\end{abstract}

\begin{IEEEkeywords}
Diffraction, wireless channel modeling, maximum likelihood estimation, shape estimation, millimeter-wave sensing, THz sensing, integrated sensing and communication (ISAC).
\end{IEEEkeywords}
\maketitle

\section{Introduction}\label{sec:intro}
Conventional wireless communication channel models either (i) assume unobstructed
line-of-sight (LoS) paths or (ii) capture multipath through specular
reflections and diffuse scattering. These abstractions neglect the
rich diffraction phenomena produced by obstacles whose transverse
dimension is on the order of a few wavelengths or more.  Ignoring diffraction not only negatively affects coverage prediction, but also discards valuable geometric information that could be exploited for integrated sensing and communication (ISAC).
When an object partially blocks the receive aperture, the \textcolor{black}{Huygens--Fresnel} principle predicts a gradual field transition, yielding a spatial response that encodes the object’s shape, range, and orientation. These signatures can be leveraged for sensing without wideband radar waveforms or stringent transmitter-receiver synchronization.

Diffraction modeling has a long history in classical electromagnetics, including the Fresnel and Fraunhofer integrals for diffraction field approximations and other methods from the Geometrical Theory of Diffraction (GTD)~\cite{6645140}. However, the models typically used for wireless communications systems ignore such effects. Recent advances in ISAC, millimeter wave (mmWave) and THz sensing have begun to explore electromagnetic (EM)-based modeling of the radio environment, particularly for reflection-dominant scenarios~\cite{sambon2025electromagnetic,10538217,10542343,Ziganshin26}. Less work has concentrated on diffraction-dominant settings, which present unique opportunities and challenges. While diffraction typically reduces signal power, it provides geometric information about objects encountered by propagating signals. Prior studies on wedge diffraction~\cite{6645140} form the basis for edge-based channel modeling, yet do not fully address shape estimation, especially in near-field scenarios. 
In parallel, shape and direction-of-arrival (DoA) estimation have been extensively studied in radar and sonar~\cite{17564}, focusing on strong scattering or reflection points. \textcolor{black}{Our contribution is the embedding of source and object parameters into a common complex array manifold, so that established array-processing techniques and information bounds can be applied to their coupled estimation.} 

\subsection{Related Work}
\label{sec:related}

\subsubsection{Antenna Array Signal Processing}
Classical array signal processing has focused on beamforming and source localization using, for example, the {Minimum-Variance Distortionless-Response} (MVDR) approach~\cite{1449208} and subspace-based methods such as {Multiple Signal Classification} (MUSIC)~\cite{1143830}. Many of these methods can be cast in a subspace-fitting framework that, with proper weighting, is asymptotically equivalent to maximum likelihood (ML) estimation and \textcolor{black}{yields} performance that asymptotically achieves the Cram\'er--Rao bound (CRB) \cite{80966,120802,348129}. This prior work has assumed unobstructed Radio-Frequency (RF) propagation from the signal sources to the receive antenna array, involving models that only focus on signal parameters such as amplitude, DoA, range, polarization, etc., rather than the propagation environment itself. Instead, in this work we study diffraction-aware ML estimation of not only the signal parameters, but also parameters associated with objects in the environment that partially block the signals before reaching the array, taking the results of \cite{80966,120802,348129} to a new physical regime and bridging traditional array processing and emerging wireless sensing applications.

\subsubsection{Near-Field Communications and Channel Modeling}
The move of communication systems to higher frequency bands has led to consideration of extremely large aperture arrays~\cite{9903389}, whose radiating near-field region is large enough to encompass objects and communication users~\cite{10220205}. Studies have developed non-uniform spherical wave and Green's function based models~\cite{10192541} for LoS channels in such scenarios, as well as physical propagation- or correlation-based stochastic models for non-LoS (NLoS) channels~\cite{10716601}.  Simpler non-EM-oriented approaches have been proposed that account for blockages using shadow or mask-based ``visibility regions'' for the receive array, and algorithms have been developed for joint channel and shadow estimation \cite{de2020non,HanYu23,10509715,Huiping25}. However, except for that noted in the subsection below, relatively little work from the communications literature has focused on modeling reflecting/diffracting objects located within the near field of a receiver array.

\subsubsection{EM-Based Blockage Modeling in Wireless Channels}
Accurate blockage models require linking rigorous EM diffraction theory with tractable wireless-channel abstractions. A knife-edge diffraction (KED) model was used in \cite{Ceulemans25} to go beyond a simple binary-mask model for blockages, but the diffraction pattern was not used to sense the blockage. The influential {Uniform Geometrical Theory of Diffraction} (UTD) provides a high-frequency model with closed-form expressions valid across reflection and shadow boundaries \cite{1451581}. In \cite{9997537}, the authors compared KED, UTD, and physical-optics predictions against wireless channel data for randomly oriented blockers, finding that KED over-estimates loss when multiple curved edges interact.
%Controlled measurements at 60–300 GHz have revealed significant attenuation around small objects such as corners, cylinders, and human bodies \cite{6134693}.  
\textcolor{black}{Many wireless-channel studies use UTD/KED loss terms or empirical attenuation models for link prediction. Such models can also support inverse sensing when appropriately parameterized, but the measurements and parameters differ from those considered here.} In contrast, the blockage models we consider are based on Huygens' principle and the Fresnel diffraction integral, which interpret every point on a wavefront as a secondary radiator whose spherical wavelets superimpose to form the next wavefront, providing a consistent approach for both near- and far-field propagation analysis.

\subsubsection{Diffraction‐Based Imaging}
Unlike specular \textcolor{black}{reflections, which} require a specific LoS geometry, edge diffraction can illuminate objects in deep \textcolor{black}{shadow and} enable around-corner sensing. Recent work leverages these properties for high-resolution imaging, user localization, and ISAC. Early work on diffraction-based imaging focused on human-centric sensing. In \cite{7881087}, the authors used a double KED model to explain human blockages at 73 GHz and showed that diffracted energy contains stable geometric cues even when the LoS is fully obstructed. Diffraction can also assist positioning when strong reflections are absent, such as in \cite{10879253} which leveraged the GTD and introduced an innovative approach for NLoS positioning.

%%%%%%%%%

%Fifth–generation (5G) and upcoming 6G systems rely increasingly on millimeter-wave (mmWave) and terahertz (THz) spectral windows to sustain multi-gigabit data rates and ultra-low latency \cite{8732419}.
%
%Accurate modeling of propagation channels is fundamental in wireless communications, radar sensing, and signal processing, as it directly impacts system design and parameter estimation. Traditional channel models often assume unobstructed paths or incorporate multipath propagation, capturing reflections and scattering phenomena in the environment. 
%
%However, at shorter wavelengths, electromagnetic waves are readily
%blocked by everyday objects—smartphones, human limbs, foliage, or
%window frames—leading to abrupt signal fading and severe coverage
%holes. Therefore, obstacles within the near-field region of the receive array can partially block the array, introducing diffraction effects that significantly alter the characteristics of the received signal~\cite{10117500}.

\subsubsection{Wave-Based Inverse Diffraction and Device-Free Sensing}
\textcolor{black}{Rampa \emph{et al.} have developed parametric diffraction-based device-free localization and accuracy bounds from RF measurements~\textcolor{black}{\cite{rampa2015physical}}, extended scalar models beyond the paraxial knife-edge construction~\textcolor{black}{\cite{rampa2017em}}, and used an EM-inspired link model~\textcolor{black}{\cite{rampa2022multiple}}. Their subsequent antenna-array study also describes diffraction-based multi-link signatures~\textcolor{black}{\cite{rampa2022arrays}}. The distinction with our work is the explicit joint source/environment spatial manifold, its ML-based treatment and CRB analysis assuming unknown communication signals, and low-dimensional shape/range/material estimation from complex array observations.}

The present work is also connected to a broader literature on
wave-based inverse imaging. Classical scalar diffraction theory and
Fourier optics provide the \textcolor{black}{Huygens--Fresnel}, Fresnel, and Fraunhofer
propagation models that underlie our diffraction-channel formulation
\cite{born1999principles,goodman2017fourier}, and inverse diffraction
problems have long been studied in optics through phase retrieval
\cite{gerchberg1972practical,fienup1982phase,teague1983deterministic,shechtman2015phase},
diffraction tomography, and inverse EM
scattering, which recover object structure, support, or material
contrast from measured scattered fields
\cite{wolf1969three,devaney1982filtered,colton2019inverse,pastorino2010microwave}.
The fundamental idea behind the GTD-based approach in \cite{11173623} is similar to that explored in this paper, but a non-parametric scheme is employed to reconstruct the environmental scene point-by-point, \textcolor{black}{which requires many more unknowns and measurements than the low-dimensional model considered here.} These works demonstrate that diffraction patterns contain recoverable geometric and phase information. However, they address optical or tomographic imaging settings and find dense field or contrast maps, while we consider scenarios where a low-dimensional parametric model is sufficient for characterizing blockage-induced diffraction. \textcolor{black}{Our representation makes joint estimation of source and blockage parameters possible using array measurements.}

A body of complementary research exploits blockage effects for device-free sensing. In these approaches, radio tomographic imaging is used to construct attenuation maps from received-signal-strength (RSS) measurements across many links
\cite{wilson2010radio,patwari2010rf}, in some cases with diffraction-inspired measurement models \cite{wang2015diffraction}.
In addition, optical and RF NLoS systems recover hidden scenes or
detect, track, and localize targets around corners from indirect
propagation paths \cite{faccio2020nlos,scheiner2020seeing,yue2022cornerradar}.
These studies reinforce the principle that non-specular wave phenomena provide useful scene information, but they generally rely on scalar RSS measurements and link-level attenuation models. \textcolor{black}{Our emphasis is on complex array-field representations, with both the illuminating-source and the object parameters entering the same manifold, unlike RSS/link attenuation fitting and image-domain inversion.}
\vspace{-0.1cm}

\subsection{Motivations and Contributions}

\begin{table*}[!t]
\caption{\textcolor{black}{Representative sensing approaches, organized by measurement domain and environmental representation.}}
\label{tab:my-table}
\centering\small
\begin{tabular}{|p{0.19\linewidth}|p{0.34\linewidth}|p{0.37\linewidth}|}
\hline
\textbf{Measurement domain} & \textbf{Environmental representation and references} & \textbf{Parameters and distinguishing features} \\ \hline
RSS or link attenuation & Radio tomography~\cite{wilson2010radio,patwari2010rf}; diffraction-based device-free localization and multi-body link models~\textcolor{black}{\cite{rampa2015physical,rampa2017em,rampa2022multiple}} & \textcolor{black}{Object position, size, or orientation, inferred from attenuation over one or more wireless links; complex per-element phases generally not used.} \\ \hline
Complex array fields & Array diffraction/body models~\textcolor{black}{\cite{rampa2022arrays}}; visibility-region and shadow-aware arrays~\cite{de2020non,HanYu23}; \textcolor{black}{joint parametric source/environment ML (this work)} & \textcolor{black}{Joint source and low-order object range, shape, material, thickness and tilt estimation from a diffraction-dependent array manifold using ML and CRBs.} \\ \hline
Image-domain reconstruction & RF tomography~\cite{wilson2010radio}; inverse diffraction tomography~\cite{wolf1969three,devaney1982filtered}; diffraction-assisted imaging~\cite{11173623} & \textcolor{black}{Dense spatial reflectivity, attenuation, or contrast maps rather than a low-dimensional array manifold.} \\ \hline
\end{tabular}
\end{table*}

\textcolor{black}{Most ISAC architectures sense through specular returns from
an illuminated target, estimating delay and angle using echo measurements.
While effective, such systems depend on
sufficient signaling bandwidth, a timing reference, or dedicated sensing
resources~\cite{9737357}.  Diffraction-based sensing takes a different approach, asking what can be inferred about an intervening object from the
spatially structured \emph{forward} field received at an array. The object need not generate an isolated, detectable specular
echo; signals of opportunity may illuminate it, and the receive array
can exploit the induced complex spatial response to estimate its geometry.
This sensing mechanism can reduce reliance on
wideband delay estimation in suitable configurations.}

\textcolor{black}{A low-dimensional scene description is attractive when a dense pixel-domain reconstruction is unnecessary. While a binary
visibility region can capture coarse shadowing, it removes the Fresnel
transition and phase oscillations that help distinguish nearby object
shapes and ranges.  Conventional KED/UTD-based link-budget models offer
improved attenuation prediction in some scenarios, but a scalar
loss coefficient alone generally does not reproduce the complex
array-wide diffraction structure used here.  Conversely, a dense
Maxwell-based reconstruction may entail substantially more unknowns
and repeated full-wave solutions than are warranted when the desired
information is just a small set of dimensions and material properties.
The central question is therefore whether one can retain enough of the
wave physics in a low-dimensional array manifold to obtain useful
estimation accuracy with practical complexity.}

\textcolor{black}{The distinction among \emph{scene representation},
\emph{measurement domain}, and \emph{propagation model} is key.
Table~\ref{tab:my-table} organizes representative prior methods by
the signals used in the inverse problem and the form of their
unknown environmental parameters.  Link/RSS models including \cite{rampa2015physical,rampa2017em,rampa2022multiple} 
already demonstrate parametric
wave-based device-free localization; image-domain methods reconstruct
much higher-dimensional spatial fields or contrast maps; and some
complex-array approaches retain spatial phase but model the blockage
primarily through visibility masks.  Our objective is to bring \emph{jointly
parameterized illumination sources and environmental objects} into the
same received-array manifold, so that ML estimation, CRBs and model-order testing can be carried out
using the statistical tools of array signal processing.}

\iffalse
\textcolor{black}{A useful physical organizing parameter is the Fresnel
number $F=d^2/(\lambda R)$ for object transverse dimension $d$ and
object-to-array distance $R$.  Two scenes with geometries and
wavelengths scaled in a way that preserves the dimensionless
Fresnel coordinates can have similar \emph{normalized} diffraction
patterns.  This connects the numerical examples to a wider range of
carrier frequencies and provides a basis for choosing the receive
aperture and element sampling; see Section~\ref{subsec:scaling}.
For near-field illumination, source wavefront curvature introduces
an additional distance scale, so object range and source range may
both affect these dimensionless signatures.}
\fi

\textcolor{black}{The contributions of the current work are the following:}
\begin{itemize}
\item \textcolor{black}{\emph{Unified source--environment manifold:} A differentiable array response with far- or near-field illumination and low-order environmental representations, including objects with finite thickness, an obliquity factor, and effective dielectric transmission.}
\item \textcolor{black}{\emph{Coupled estimation and limits:} Stochastic ML formulations for source and object parameter estimation, explicit covariance/noise profiling, and nuisance-parameter-adjusted CRBs that diagnose local observability, including potential shape--range and thickness--material coupling.}
\item \textcolor{black}{\emph{Order selection and model comparisons:} A minimum description length (MDL)-type search for the number of sources and object parameter dimensions, and simulations for different shape parameterizations with blockage-aware and blockage-free reference models.}
\end{itemize}
\textcolor{black}{A preliminary version of this work appeared in~\cite{10942642}.}

\iffalse
\textcolor{black}{Two practical extensions make the framework more than an
opaque-mask shape fit.  First, the environmental parameter vector can
include nonzero axial thickness and an effective complex dielectric
response, although the coupling between thickness and permittivity
can be severe unless the measurement configuration provides adequate
frequency or angular diversity. Second, the number of polygon
vertices need not be fixed a priori: a source-count MDL statistic
and a penalized profile-likelihood criterion provide an empirical
route to selecting the number of sources and the polygon complexity.
The simulation section deliberately distinguishes studies in which
source positions are jointly unknown from studies in which the known
illumination directions isolate the environmental parameters.}
\fi

\subsection{Paper Organization}

The remainder of the paper is organized as follows. 
Section~\ref{sec:model} introduces the general diffraction channel
model, while Section~\ref{sec:analytical} discusses several closed-form results for blockages with canonical shapes. 
\textcolor{black}{Section~\ref{sec:Shape} presents the shape representations and joint source-localization/environmental estimation methods, including model-order selection.}
Section~\ref{sec:CRB} briefly derives the Cram\'er--Rao bounds and discusses some identifiability results. Simulation examples are provided in Section~\ref{sec:sim}, and conclusions and future directions follow in Section~\ref{sec:conclusion}.
%%%%%%%

\section{Physically-Consistent Diffraction Channel Modeling}
\label{sec:model}

\begin{figure}[t!]
    \begin{center}
        \includegraphics[scale=0.35]{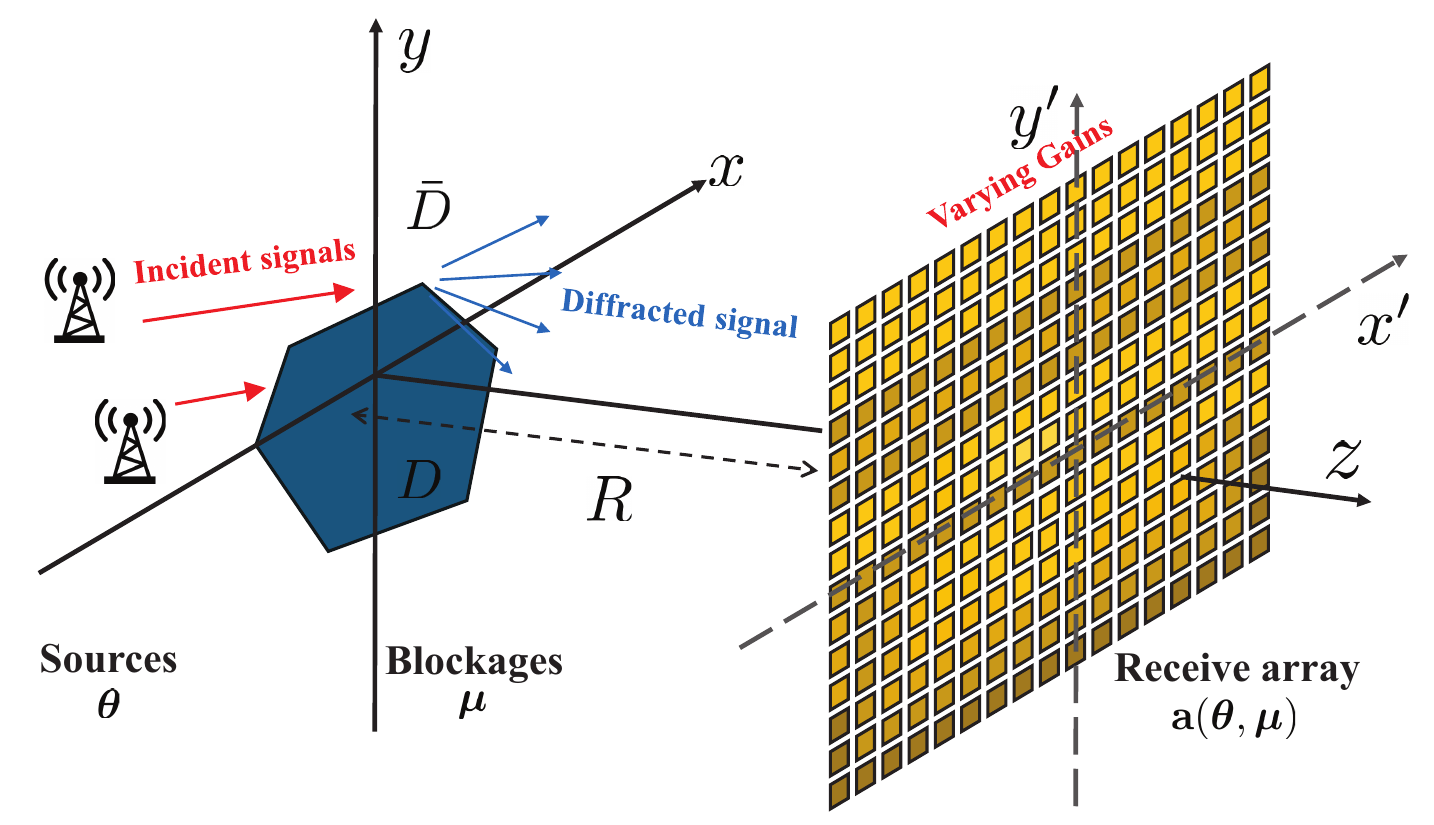}
        \caption{Illustration of the considered scenario.}
        \label{Fig:cover}
    \end{center}
\end{figure}

In this section, we derive a diffraction-based array response model
\textcolor{black}{for the general scenario in Fig.~\ref{Fig:cover}, in which sources may be in either the far field or near field of the receive array and an intervening object modifies the incident field.} The blockage does not simply create a binary visibility region at the 
array, but instead generates a field with a structured
diffraction pattern whose spatial variations depend on the object
shape, range, and the source directions. \textcolor{black}{We use low-dimensional object descriptions and assume their identifiability from the resulting receive-array signatures.} Throughout the paper, we denote the wavenumber by $k_0=2\pi/\lambda$ and absorb the global phase and amplitude factors common to all receive antennas into the source waveform. 

\subsection{Aperture-Mask and Scalar Diffraction Model}

As shown in Fig.~\ref{Fig:cover}, consider a planar object located in
the plane $z=0$, parallel to the receive array, which lies in the plane
$z=R$. A point on the object
plane is denoted by
\[
    \mathbf r_s(x,y) = [x,y,0]^T,
\]
and the position of the $m$th receive antenna is
\[
    \mathbf r_m = [x_m',y_m',R]^T .
\]
Let $D(\boldsymbol\mu)\subset \mathbb R^2$ denote the opaque
cross-section of the blockage in the $x$-$y$ plane, where ${\boldsymbol{\mu}}$ denotes the set of parameters used to describe the physical characteristics of the blockage (to be described in detail later). We define the indicator function of the blockage by
\[
    \chi_D(x,y;\boldsymbol\mu)
    =
    \begin{cases}
    1, & (x,y)\in D(\boldsymbol\mu),\\
    0, & (x,y)\notin D(\boldsymbol\mu),
    \end{cases}
\]
and the corresponding aperture transmission function by
\[
    \tau(x,y;\boldsymbol\mu)
    =
    1-\chi_D(x,y;\boldsymbol\mu).
\]
Thus, $\tau=0$ inside the blockage and $\tau=1$ in the open region.
This convention will be used throughout the paper.

\textcolor{black}{Take the
reference antenna location to be $\mathbf r_{\rm ref}=[0,0,R]^T$,
and let the $k$th source be described by}
\begin{equation}\label{eq:source_vector_nf_expanded}
\textcolor{black}{\boldsymbol\theta_k=
[\varphi_k,\,\eta_k,\,d_k]^T,\qquad
\boldsymbol\theta=[\boldsymbol\theta_1^T,\ldots,
\boldsymbol\theta_K^T]^T,}
\end{equation}
\textcolor{black}{where $\varphi_k$ and $\eta_k$ are the source azimuth and
elevation, and $d_k>0$ is its distance to the array reference.
For a far-field source, $d_k$ is not identifiable from a plane-wave
array manifold and the source parameter vector reduces to the two
angles.  The unit vector pointing in the direction of
wave propagation toward the array is}
\begin{equation}\label{eq:source_direction_expanded}
\textcolor{black}{\mathbf q_k =
[\sin\varphi_k\cos\eta_k,\,\sin\eta_k,\,
\cos\varphi_k\cos\eta_k]^T.}
\end{equation}
\textcolor{black}{The source position is $\mathbf p_k=\mathbf r_{\rm ref}-d_k\mathbf q_k$, and
we assume its $z$-coordinate is negative, so it lies on the
illumination side of the object plane.  Equivalently, we may write
$\mathbf p_k=[p_{x,k},p_{y,k},-L_k]^T$, where
$L_k=d_k q_{z,k}-R>0$ is the source-to-object-plane axial
separation.  The scalar point-source illumination on that plane is}
\begin{equation}\label{eq:nf_incident_new}
\textcolor{black}{U_k^{\rm inc}(x,y)=\alpha_k
\frac{\exp\!\left(jk_0\|\mathbf r_s(x,y)-\mathbf p_k\|\right)}
{\|\mathbf r_s(x,y)-\mathbf p_k\|}.}
\end{equation}
\textcolor{black}{The complex source-dependent coefficient $\alpha_k$
includes arbitrary reference phase and gain values.
As $d_k\rightarrow\infty$ while $\mathbf q_k$ remains fixed,
the normalized incident field is proportional to a plane wave:}
\begin{equation}\label{eq:ff_incident_expanded}
\textcolor{black}{U_k^{\rm inc}(x,y)\ \propto\
\exp\!\left[jk_0(q_{x,k}x+q_{y,k}y)\right].}
\end{equation}
\textcolor{black}{Thus, the same diffraction operator accommodates both
near-field source localization and the
far-field special case.}

Under the scalar \textcolor{black}{Huygens--Fresnel} approximation, the diffracted field
at the $m$th receive antenna due to the $k$th source is
\begin{align}
    U_{m,k}
    &=
    \iint_{\mathbb R^2}
        \tau(x,y;\boldsymbol\mu)
        U_k^{\rm inc}(x,y)
        K(\mathbf r_m,\mathbf r_s)
    \,dx\,dy,
    \label{eq:diff_op}
\end{align}
where
\begin{align}
    K(\mathbf r_m,\mathbf r_s)
    &=
    \frac{1}{j\lambda}
    \frac{\exp\left(jk_0\rho_m(x,y)\right)}
         {\rho_m(x,y)}
    \mathcal O_m(x,y),
    \label{eq:RS_kernel} \\
    \rho_m(x,y)
    &=
    \|\mathbf r_m-\mathbf r_s(x,y)\| \\
    &=
    \sqrt{(x_m'-x)^2+(y_m'-y)^2+R^2}, 
\end{align}
\textcolor{black}{and the obliquity factor is $\mathcal O_m(x,y)=R/\rho_m(x,y)=\cos\psi_m$. The first Rayleigh--Sommerfeld kernel additionally contains a relative correction $1+j/(k_0\rho_m)$ that is negligible when $k_0\rho_m\gg1$. Including this factor in \eqref{eq:RS_kernel} does not alter the subsequent paraxial formulas. We refer to \eqref{eq:diff_op}--\eqref{eq:RS_kernel} as the nonparaxial scalar model. 
%The actual scalar path length, spreading, and obliquity are retained, but Maxwell's vector boundary-value problem is not solved. 
In the Fresnel approximation, $\mathcal O_m\simeq1$ and $1/\rho_m\simeq1/R$. The scalar model \eqref{eq:diff_op} is adopted under the following
conditions: (i) the object can be described by a scalar transmission screen or a thin-slice dielectric approximation; paraxiality is needed only for the Fresnel simplification, not for the general distance-dependent kernel; (ii) the axial thickness of the object is small compared with the range $R$, so that the object can be represented by a single transmission plane; this is separate from the transverse dimension, which determines the Fresnel number, and from the receive aperture, which determines whether the paraxial form \eqref{eq:fresnel_mask_field} applies; (iii) polarization effects and edge (fringe) currents are neglected, reasonable for electrically large, strongly attenuating objects, but possibly inaccurate for dielectric refraction, polarization conversion, and metallic edge fields~\cite{8883197,9997537}; and (iv) each data block is narrowband, but several separate frequencies may be used jointly.} When these conditions
do not hold, more accurate scalar or vector diffraction models (e.g., UTD or
physical optics) or full-wave solvers must replace the kernel
\eqref{eq:RS_kernel}. The estimation framework developed in 
Sections~\ref{sec:Shape} and \ref{sec:CRB} can still be used in such cases, since it requires only a differentiable forward model
$\mathbf a(\bm\theta,\bm\mu)$.

\textcolor{black}{Let $U_{m,k}^{(0)}$ be the field computed with no blockage ($\tau\equiv1$), using the same source and propagation geometry as the blocked response, and define $g_{m,k}(\boldsymbol\mu)=U_{m,k}/U_{m,k}^{(0)}$. An unobstructed source has $g_{m,k}=1$ regardless of its range. If we let ${\mathbf a}^{(0)}(\boldsymbol\theta_k)$ represent the array manifold vector for source $k$ without a blockage, then the corresponding array manifold with the blockage is $\mathbf{a}(\boldsymbol{\theta}_k,\boldsymbol{\mu})={\mathbf a}^{(0)}(\boldsymbol\theta_k) \odot {\mathbf g}_k(\boldsymbol\mu)$, where ${\mathbf{g}}_k(\boldsymbol\mu)=[g_{1}(\boldsymbol\theta_k,\boldsymbol\mu) \; \cdots \; g_{M}(\boldsymbol\theta_k,\boldsymbol\mu)]^T$.} The corresponding array response matrix and received signal model are
\begin{align}
    \mathbf A(\boldsymbol\theta,\boldsymbol\mu)
    &=
    \left[
        \mathbf a(\boldsymbol\theta_1,\boldsymbol\mu),
        \ldots,
        \mathbf a(\boldsymbol\theta_K,\boldsymbol\mu)
    \right] \\
    \mathbf s(t)
    &=
    \mathbf A(\boldsymbol\theta,\boldsymbol\mu)\mathbf x(t)
    +
    \mathbf n(t),
    \label{receive signal}
\end{align}
where $\mathbf x(t)$ contains the complex source signals, including
source-dependent propagation gains and common phase factors, and
$\mathbf n(t)$ is additive noise.

\subsection{\textcolor{black}{Extension to Objects with Finite Thickness}}

\textcolor{black}{For an object with non-negligible thickness $h$, we replace the thin opaque mask model with a scalar split-step description. Let $U_\ell(x,y)$ denote the field at the entrance of layer $\ell$ of $N_z$ layers, each of depth $\Delta z=h/N_z$. Let $\chi_\ell(x,y)\in\{0,1\}$ indicate whether the transverse position $(x,y)$ lies inside ($=1$) or outside ($=0$) the object in slice $\ell$. We use the local-wavenumber update}
\begin{equation}\label{eq:splitstep_new}
\textcolor{black}{U_{\ell+1}= (1-\chi_\ell)\mathcal P_{\Delta z}^{(1)}\{U_\ell\}
+\chi_\ell\mathcal P_{\Delta z}^{(n_r)}\{U_\ell\},}
\end{equation}
\textcolor{black}{where $n_r$ is the complex refractive index of the material and the scalar angular-spectrum propagator is}
\begin{equation}\label{eq:splitstep_angular_spectrum}
\textcolor{black}{\mathcal P_{\Delta z}^{(n)}\{U\}
=\mathcal F_{\perp}^{-1}\!\left\{
\exp\!\left[j\Delta z\sqrt{k_0^2n^2-\|\mathbf k_\perp\|^2}\right]
\mathcal F_\perp\{U\}\right\}.}
\end{equation}
\textcolor{black}{Thus, each layer is propagated with the material wavenumber inside the object and the free-space wavenumber outside it. The square-root branch is chosen so that evanescent components decay with propagation. After the final layer, the field propagates in free space to the receiving array. Fractional coverage of numerical cells can be used at the material boundaries. In the example presented in Section~VI-G, the propagation acts on one transverse coordinate and the computed object-induced field is added to the analytically known unblocked cylindrical-wave response. This decomposition avoids artificial FFT wrap-around of the incident field.}

\textcolor{black}{The longitudinal-wavenumber propagator in~\eqref{eq:splitstep_angular_spectrum} accounts for the dependence of phase and attenuation on the angle of propagation inside the dielectric. This is useful for obliquely incident fields. The local-wavenumber method is a forward-only scalar approximation, and does not enforce all vector EM boundary conditions or include all reflections and multiple scattering. Thus, it offers a lower-complexity alternative to a full EM boundary-value calculation. For approximately plane-parallel dielectric objects, another option is a coherent, angle-dependent slab transmission coefficient, as used in the superellipse example of Section~VI-E. Tilted objects can be represented using geometry-dependent propagation distances and local incidence angles.}

\textcolor{black}{The environmental vector $\bm\mu$ may therefore include object range, thickness, projected shape, the real relative permittivity $\varepsilon_r'$, dielectric loss tangent $\tan\delta$, and object tilt. These parameters may be strongly coupled; for example, thickness and dielectric contrast can give similar phase shifts and attenuation. Multiple frequencies and illumination directions can improve the conditioning of their estimation, as illustrated later. The thin opaque-screen model is still useful for inexpensive analytic insight and initialization, but is most appropriate when the transmitted field is negligible (for example, $k_0\operatorname{Im}\{n_r\}h\gg1$ for a lossy dielectric or several skin depths for a conductor) and the object spans several wavelengths transversely.}

\subsection{\textcolor{black}{Paraxial Fresnel Model and Babinet Normalization}}

When the lateral distances satisfy
\[
    |x_m'-x|\ll R,
    \qquad
    |y_m'-y|\ll R,
\]
the propagation distance can be approximated by
\begin{equation}
    \rho_m(x,y)
    \simeq
    R+
    \frac{(x_m'-x)^2+(y_m'-y)^2}{2R}.
    \label{eq:fresnel_distance}
\end{equation}
Substituting \eqref{eq:fresnel_distance} in
\eqref{eq:diff_op} gives the Fresnel propagation formula
\begin{align}
    U_{m,k}
    &\simeq
    \frac{\alpha_k e^{jk_0R}}{j\lambda R}
    \iint_{\mathbb R^2}
        \tau(x,y;\boldsymbol\mu)
        \exp\left\{
            j k_0(q_{x,k}x+q_{y,k}y)
        \right\}
        \nonumber\\
    &\quad\times
        \exp\left\{
            j\frac{k_0}{2R}
            \left[
                (x_m'-x)^2+(y_m'-y)^2
            \right]
        \right\}
    \,dx\,dy .
    \label{eq:fresnel_mask_field}
\end{align}
Define the source-dependent shifted receive coordinates
\[
    \widetilde x_{m,k}=x_m'-R q_{x,k},
    \qquad
    \widetilde y_{m,k}=y_m'-R q_{y,k},
\]
so that
\begin{align}
    a_{m}^{(0)}(\boldsymbol\theta_k)
    &=
    \exp\left\{
        j k_0(q_{x,k}x_m' + q_{y,k}y_m')
    \right\} \\
    a_m(\boldsymbol\theta_k,\boldsymbol\mu)
    &=
    a_{m}^{(0)}(\boldsymbol\theta_k)
    g_m(\boldsymbol\theta_k,\boldsymbol\mu),
    \label{a}
\end{align}
where
{
\begin{align}
    g_m(\boldsymbol\theta_k,\boldsymbol\mu)
    = 1 - \frac{1}{j\lambda R} & \iint_{D(\boldsymbol\mu)}
        \exp\left\{j\frac{k_0}{2R}
            \Bigl[ (x-\widetilde x_{m,k})^2 \right. \Bigr. \nonumber \\
                &\quad\biggl. \Bigl.
                + (y-\widetilde y_{m,k})^2
            \Bigr]
        \biggr\}
    \,dx\,dy .
    \label{eq:babinet_fresnel_physical}
\end{align}}
{\flushleft Eq.~\eqref{eq:babinet_fresnel_physical}} \textcolor{black}{is the computational form of
Babinet's principle,} which states that the field behind an opaque object is the
unobstructed field minus the field that would pass through an aperture with the same \textcolor{black}{support $D(\boldsymbol\mu)$. Integrating over the bounded region $D$ instead of the unbounded open region $\bar D$ is what makes numerical evaluation practical, and it enforces the normalization $g_m=1$ when no blockage is present ($D=\emptyset$). The source DoAs enter \eqref{eq:babinet_fresnel_physical} only through the shifted coordinates $\widetilde x_{m,k}$ and $\widetilde y_{m,k}$, so the diffraction pattern depends on the DoAs not only through the conventional steering phase $a_m^{(0)}$, but also through a lateral displacement of the Fresnel pattern across the receive aperture.}
%Thus, in the absence of a blockage, the response vector reduces to the ordinary plane-wave response vector.

We now introduce dimensionless Fresnel coordinates
\[
    u=\sqrt{\frac{2}{\lambda R}}\,x,
    \qquad
    v=\sqrt{\frac{2}{\lambda R}}\,y,
\]
and the shifted receive coordinates
\[
    u_{m,k}'=
    \sqrt{\frac{2}{\lambda R}}\,
    \widetilde x_{m,k},
    \qquad
    v_{m,k}'=
    \sqrt{\frac{2}{\lambda R}}\,
    \widetilde y_{m,k}.
\]
Let $D_s(\boldsymbol\mu,R)$ denote the scaled version of the blockage:
\[
    D_s(\boldsymbol\mu,R)
    =
    \left\{
        (u,v):
        \left(
            \sqrt{\frac{\lambda R}{2}}u,
            \sqrt{\frac{\lambda R}{2}}v
        \right)
        \in D(\boldsymbol\mu)
    \right\}.
\]
Then \eqref{eq:babinet_fresnel_physical} becomes
\begin{align}
    g_m(\boldsymbol\theta_k,\boldsymbol\mu)
    =
    1
    -
    \frac{1}{2j}
    & \iint_{D_s(\boldsymbol\mu,R)}
        \exp\left\{
            j\frac{\pi}{2}
            \Bigl[
                (u-u_{m,k}')^2 \Bigr. \right. \nonumber \\
 & \quad\Bigl. \Bigl. +(v-v_{m,k}')^2
            \Bigr]
        \Bigr\}
    \,du\,dv .
    \label{eq:babinet_fresnel_scaled}
\end{align}
{\flushleft{Equivalently,}} \textcolor{black}{$g_m$ is $1/(2j)$ times the same integral of $\tau_s=1-\chi_{D_s}$ over $\mathbb R^2$, since the kernel integrates to $2j$ over the plane.}
%The finite-domain expression in \eqref{eq:babinet_fresnel_scaled} is usually more convenient numerically.

\subsection{\textcolor{black}{Paraxial Spherical-Wave Illumination}}
\textcolor{black}{A useful near-field simplification follows by retaining quadratic phase for both illumination and propagation. If a source is at transverse position $\mathbf p_{\perp,k}$ and axial distance $L_k$ from the object plane, set $\mathbf r_{\perp,m}=(x_m',y_m')^T$ and define}
\begin{equation}\label{eq:effective_fresnel_new}
\textcolor{black}{R_{{\rm eff},k}=\frac{L_kR}{L_k+R},\qquad
\mathbf c_{m,k}=\frac{R\mathbf p_{\perp,k}+L_k\mathbf r_{\perp,m}}{L_k+R}.}
\end{equation}
\textcolor{black}{After normalization by the unobstructed field,}
\begin{equation}\label{eq:nf_fresnel_new}
\textcolor{black}{g_{m,k}^{\rm NF}\simeq1-\frac{1}{j\lambda R_{{\rm eff},k}}
\iint_D\exp\!\left[\frac{jk_0}{2R_{{\rm eff},k}}
\|\mathbf r_\perp-\mathbf c_{m,k}\|^2\right]d\mathbf r_\perp.}
\end{equation}
\textcolor{black}{The plane-wave Fresnel scaling is recovered as $L_k\rightarrow\infty$ and $-\mathbf p_{\perp,k}/L_k$ is held equal to the source transverse direction cosine. The effective distance $R_{{\rm eff},k}$ explicitly shows how the source wavefront curvature modifies the Fresnel number and provides a low-cost near-field initialization. The nonparaxial scalar integral remains preferable for large angles.}

\subsection{Fraunhofer Approximation}

When the blockage dimensions are sufficiently small relative to the
propagation distance, the quadratic phase terms over the object support
may be neglected. More precisely, if $a_D$ is the maximum transverse
dimension of the blockage, the Fraunhofer approximation is appropriate
when
\[
    \frac{k_0 a_D^2}{2R}\ll 1.
\]
Expanding the phase in \eqref{eq:babinet_fresnel_physical} and
neglecting terms proportional to $x^2+y^2$ over
$D(\boldsymbol\mu)$ yields
\begin{align}
    g_m^{\rm FF}(\boldsymbol\theta_k,\boldsymbol\mu)
    \simeq
    1
    -
    &\frac{
        \exp\left\{
            j\frac{k_0}{2R}
            \left(
                \widetilde x_{m,k}^2
                +
                \widetilde y_{m,k}^2
            \right)
        \right\}
    }{j\lambda R} \nonumber \\
    & \qquad \times\widehat{\chi_D}
    \left(
        \frac{k_0\widetilde x_{m,k}}{R},
        \frac{k_0\widetilde y_{m,k}}{R}
    \right),
    \label{eq:fraunhofer_blockage}
\end{align}
where
\[
    \widehat{\chi_D}(\kappa_x,\kappa_y)
    =
    \iint_{D(\boldsymbol\mu)}
        e^{-j(\kappa_x x+\kappa_y y)}
    \,dx\,dy
\]
is the two-dimensional (2D) Fourier transform of the blockage indicator
function. Thus, in the far field, the diffraction distortion is
determined by the Fourier transform of the object shape. In the
Fresnel regime, however, the quadratic phase over the object cannot be
ignored, and the finite-domain Fresnel integral in
\eqref{eq:babinet_fresnel_physical} or
\eqref{eq:babinet_fresnel_scaled} must be used.

\subsection{Multiple Blockage Planes}\label{sec:multi}
The same notation also clarifies
the extension to multiple blockage planes at different ranges. Suppose that the field propagates through
$B$ planes located at $z=z_1,\ldots,z_B$, ordered from the source 
toward the receive array. Let $\tau_b(x,y)$ denote the transmission
function of the blockage in the $b$th plane. The field is
updated by alternating free-space propagation and multiplication by
the aperture mask:
\begin{align}
    U_b^-(x,y)
    &=
    \mathcal P_{\Delta z_b}
    \left\{
        U_{b-1}^+(x,y)
    \right\}
    \\
    U_b^+(x,y)
    &=
    \tau_b(x,y) U_b^-(x,y),
\end{align}
where $\mathcal P_{\Delta z_b}\{\cdot\}$ denotes Fresnel or \textcolor{black}{nonparaxial scalar} free-space propagation over the distance $\Delta z_b$ between
successive planes. The receive-array samples are obtained by one final propagation from the last blockage plane to $z=R$. \textcolor{black}{This sequential operator is the appropriate scalar model for mutually shadowing objects; in general it is not equal to a sum of isolated single-object responses. Different orders and ranges can alter both visibility and diffraction, and adding object parameters may introduce symmetries or nearly dependent Jacobian columns. Thus, global identifiability and explicit resolution conditions for overlapping objects require a substantially more detailed study, which we defer.}

\iffalse
\begin{figure}[t!]
    \begin{center}
        \includegraphics[scale=0.3]{EPSs/Diffraction_multi.pdf}
        \caption{Illustration of multiple blockages.}
        \label{Fig:multi}
    \end{center}
\end{figure}
\fi

\section{Analytical Results based on Fresnel Integrals}
\label{sec:analytical}

\begin{figure}[t!]
\centering
\subfigure[Blockage illustration.]{\label{edge1}
\includegraphics[width= 2.2in]{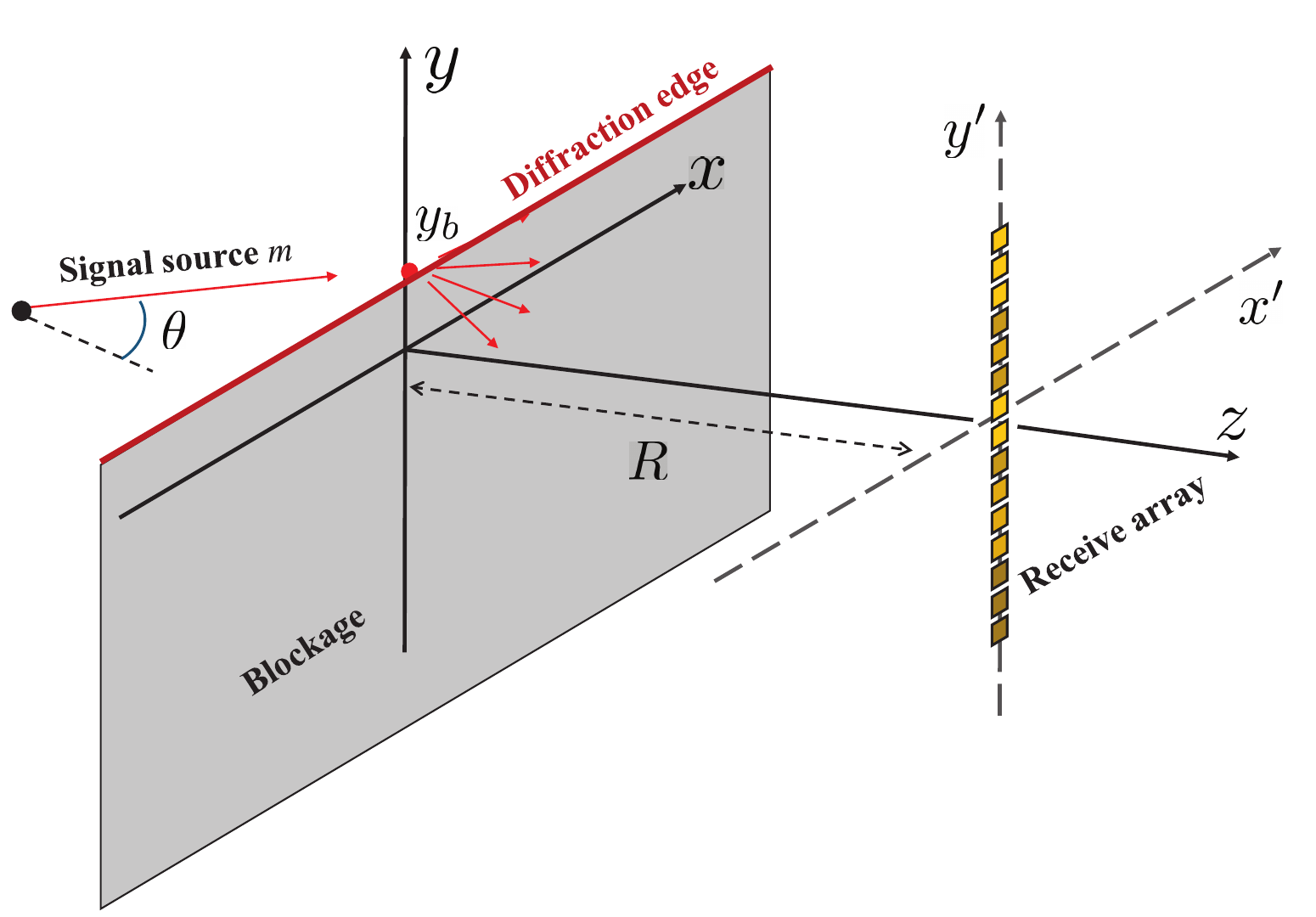}}
\subfigure[Analytical gain variation.]{\label{edge2}
\includegraphics[width= 2.2in]{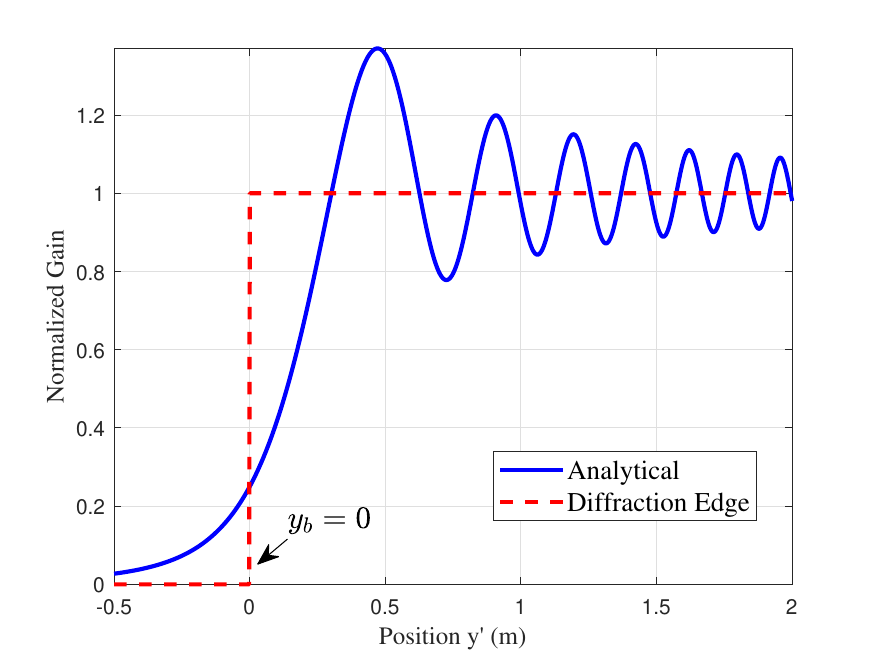}}
\caption{Diffraction pattern of a straight edge.}\label{Fig_Duo2}\label{edge:fig}%\vspace{-0.5 cm}
\end{figure}

\begin{figure}[t!]\label{strip:fig}
\centering
\subfigure[Blockage illustration.]{\label{strip1}
\includegraphics[width= 2.5in]{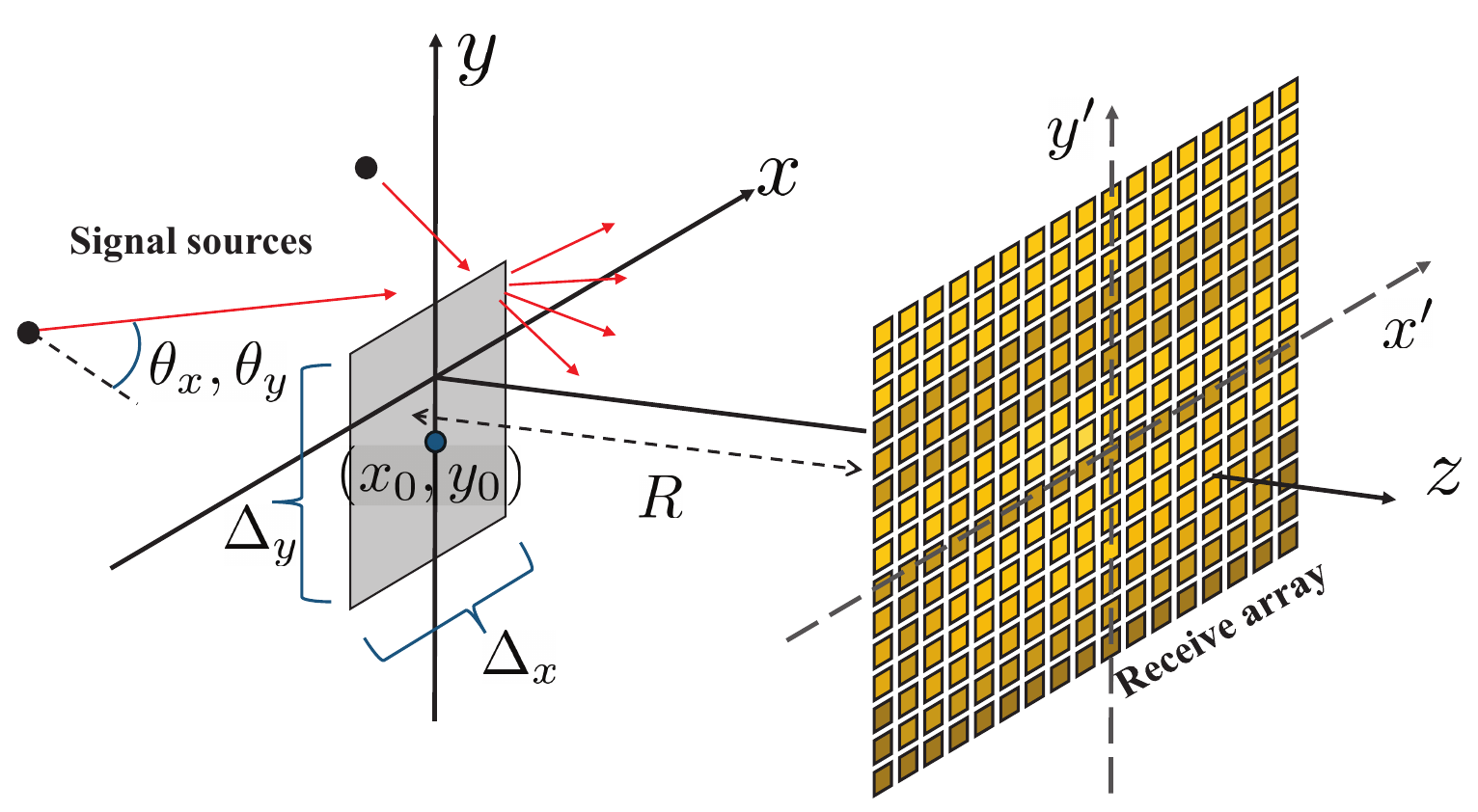}}
\subfigure[Analytical gain variation.]{\label{strip2}
\includegraphics[width= 2.5in]{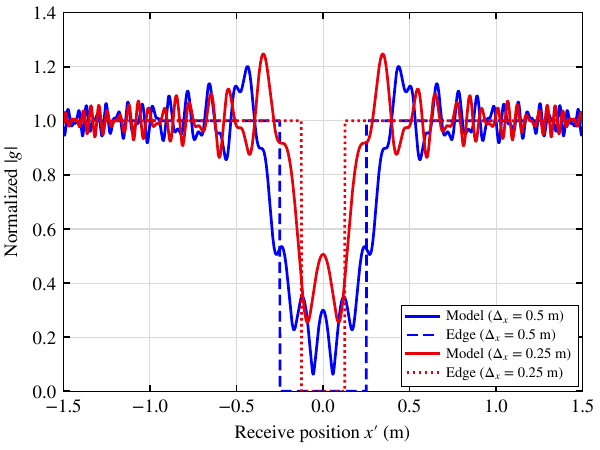}}
\caption{Diffraction pattern of a rectangle.}\label{Fig_Duo3}%\vspace{-0.5 cm}
\end{figure}

Using the Fresnel/Babinet representation in
\eqref{eq:babinet_fresnel_physical}--\eqref{eq:babinet_fresnel_scaled},
we can obtain closed-form analytical results for several canonical
blockage geometries and finite-dimensional numerical integrations for
more complicated shapes. In the following, we first present the
straight-edge model illustrated in Fig.~\ref{Fig_Duo2}, then derive the
rectangular-screen model in Fig.~\ref{Fig_Duo3}. The same
Babinet-normalized convention is used in both cases.

\subsection{Diffraction from a Straight Edge}\label{sec:edge}
Here we assume the blockage is due to a semi-infinite screen located at $z = 0$, as illustrated in Fig.~\ref{edge1}. In this case, the vector of parameters reduces to $\boldsymbol\mu=\{ y_b, R\}$, where $y_b$ is the $y$-coordinate of the diffraction edge and $R$ is the blockage range. 
The one-dimensional (1D) specialization of the diffraction factor in \eqref{a} \textcolor{black}{can be written, within the paraxial approximation, as}
\begin{equation}\label{a_edge}
    \textcolor{black}{{a}_m (y_b,R,\theta) = F\Big[\sqrt{\frac{\pi r_m}{\lambda}}\Big(\frac{y_b-y'_m}{r_m}+\sin\theta\Big)\Big]e^{jk_0y'_m\sin\theta},}
\end{equation}
where \textcolor{black}{$F(u) = \frac{e^{-j\pi/4}}{\sqrt{\pi}}\int_u^\infty e^{jv^2}\mathrm{d}v$ is the normalized complex Fresnel integral, with $F(-\infty)=1$ and $F(0)=1/2$,} $r_m = \sqrt{(y_b-y'_m)^2 + R^2}$ is the distance from the $m$th receive antenna to the edge of the blockage, and $y'_m$ is the coordinate of the $m$th receive antenna. \textcolor{black}{The sign of $\sin\theta$ is consistent with the convention in Fig.~\ref{edge1}. A positive angle has a positive transverse component in the direction of wave propagation.}
\textcolor{black}{In Fig.~\ref{edge2}, we plot the normalized received power gain $|a_m|^2$ for a straight edge at $y_b=0$ and one far-field source with $\bm{\theta} = (0,0)$ for a carrier frequency of $10$~GHz and $R=10$~m.} We see that, due to diffraction effects, the signal power gradually increases when passing the shadow edge, then oscillates around the baseline value. According to \eqref{a_edge}, the position of the peak gain and the oscillation frequency depend on $R$.

\textcolor{black}{All analytical diffraction factors $g$ are complex field-amplitude ratios, where $|g|$ denotes normalized field magnitude and $|g|^2$ normalized received power. Thus, at the geometric boundary of an ideal half-plane illuminated normally, the standard Fresnel value $g=1/2$ gives $|g|=1/2$ and $|g|^2=1/4$ (approximately $-6.02$~dB in power). In Fig.~\ref{edge2}, the normalized gain is expressed in terms of power, $|g|^2$, rather than field amplitude.}

\subsection{Diffraction \textcolor{black}{from} a Rectangular Screen}\label{sec:rect}

We next specialize
\eqref{eq:babinet_fresnel_scaled} to a rectangular opaque object, as illustrated in Fig.~\ref{strip1}. Let
the blockage be centered at $(x_c,y_c)$ with full width $\Delta_x$ and
full height $\Delta_y$, so that
\[
    D
    =
    \left[
        x_c-\frac{\Delta_x}{2},
        x_c+\frac{\Delta_x}{2}
    \right]
    \times
    \left[
        y_c-\frac{\Delta_y}{2},
        y_c+\frac{\Delta_y}{2}
    \right].
\]
In scaled Fresnel coordinates, the rectangle is
\[
    D_s
    =
    [u_1,u_2]\times[v_1,v_2],
\]
where
\begin{align}
    u_1
    &=
    \sqrt{\frac{2}{\lambda R}}
    \left(
        x_c-\frac{\Delta_x}{2}
    \right),
    &
    u_2
    &=
    \sqrt{\frac{2}{\lambda R}}
    \left(
        x_c+\frac{\Delta_x}{2}
    \right),
    \\
    v_1
    &=
    \sqrt{\frac{2}{\lambda R}}
    \left(
        y_c-\frac{\Delta_y}{2}
    \right),
    &
    v_2
    &=
    \sqrt{\frac{2}{\lambda R}}
    \left(
        y_c+\frac{\Delta_y}{2}
    \right).
\end{align}
For a source with direction cosine components $(q_x,q_y)$, the shifted
receive coordinates are
\[
    u_m'
    =
    \sqrt{\frac{2}{\lambda R}}
    (x_m'-Rq_x),
    \qquad
    v_m'
    =
    \sqrt{\frac{2}{\lambda R}}
    (y_m'-Rq_y).
\]

Substituting the rectangular support into
\eqref{eq:babinet_fresnel_scaled}, the obstruction term separates into
the product of two 1D Fresnel integrals:
\begin{align}
    g_m^{\rm rect}
    &=
    1
    -
    \frac{1}{2j}
    I_u(u_m';u_1,u_2)
    I_v(v_m';v_1,v_2),
    \label{eq:g_rect_compact}
\end{align}
where
\begin{align}
    I_u(u_m';u_1,u_2)
    &=
    \int_{u_1}^{u_2}
        \exp\left\{
            j\frac{\pi}{2}(u-u_m')^2
        \right\}
    \,du,
    \\
    I_v(v_m';v_1,v_2)
    &=
    \int_{v_1}^{v_2}
        \exp\left\{
            j\frac{\pi}{2}(v-v_m')^2
        \right\}
    \,dv.
\end{align}
For the standard Fresnel cosine and sine integrals 
\[
    C(z)=\int_0^z \cos\left(\frac{\pi}{2}t^2\right)\,dt
    \qquad
    S(z)=\int_0^z \sin\left(\frac{\pi}{2}t^2\right)\,dt ,
\]
define
\begin{align}
    C_u
    &=
    C(u_2-u_m')-C(u_1-u_m')
    \\
    C_v
    &=
    C(v_2-v_m')-C(v_1-v_m').
\end{align}
and similarly $S_u$ and $S_v$ for the sine integral. Then
\[
    I_u=C_u+jS_u,
    \qquad
    I_v=C_v+jS_v.
\]
Therefore, the normalized diffraction factor for the rectangular
blockage can be written explicitly as
\begin{align}
    g_m^{\rm rect}
    &=
    1
    -
    \frac{1}{2}
    \left(
        C_u S_v + S_u C_v
    \right)
    +
    \frac{j}{2}
    \left(
        C_u C_v - S_u S_v
    \right).
    \label{eq:g_rect_expanded}
\end{align}

The full array response for the $m$th antenna is thus
\[
    a_m(\boldsymbol\theta,\boldsymbol\mu)
    =
    \exp\left\{
        j k_0(q_x x_m' + q_y y_m')
    \right\}
    g_m^{\rm rect},
\]
up to source-dependent constants that are absorbed into the transmitted
symbol. When the blockage is absent,
$D=\emptyset$, we have $I_uI_v=0$ and hence $g_m^{\rm rect}=1$.
\textcolor{black}{Fig.~\ref{strip2} plots the normalized received field amplitude $|g|$ along the receive $x$-axis for rectangle widths $\Delta_x=0.5$ and $0.25$~m, a 10~GHz carrier, object range $R=2$~m, and normal incidence. The calculation is one-dimensional, assuming that the object extends uniformly along $y$, so $\Delta_y$ is not specified. The wider plate casts a deeper shadow with wider edge fringes, so the diffraction pattern depends strongly on the plate size.}

\section{Shape Parameterization and Estimation}
\label{sec:Shape}

For more complicated blockage shapes, generic parameterizations beyond those for straight-edge and rectangular objects are needed. In the following, we describe some standard shape parameterizations, and then present the ML estimation framework for source DoAs and blockage parameters.

\subsection{Superellipse Shape Parameterization}
For symmetrically shaped objects, the superellipse family can be used to model the blockage contour with a few generic parameters. Mathematically, the superellipse is defined as the set of points satisfying
\begin{equation}
    \biggl(\frac{|\Tilde{x}|}{a_x}\biggr)^{2n} 
    + \biggl(\frac{|\Tilde{y}|}{a_y}\biggr)^{2n}
    \;\le\;1, 
\end{equation}
where $(\Tilde{x},\Tilde{y})$ are local coordinates after translation to $(x_0,y_0)$ and rotation by $\phi$:
\begin{equation}
    \begin{pmatrix} 
      \Tilde{x} \\[6pt] 
      \Tilde{y} 
    \end{pmatrix}
    =
    \begin{pmatrix}
      \cos\phi & \sin\phi \\[4pt]
      -\sin\phi & \cos\phi
    \end{pmatrix}
    \begin{pmatrix}
      x - x_0 \\[4pt] 
      y - y_0
    \end{pmatrix}.
\end{equation}
The shape parameters to estimate for this model are thus $x_0,\,y_0$ (center coordinates), $a_x, a_y$ (semi-axis scaling),
$\phi$ (rotation angle), and $n$ (superellipse exponent). This yields a family of shapes shown in Fig.~\ref{Fig:ellipse}. 

\subsection{B-Spline Parameterization}\label{sub:bspline}
\begin{figure}[!t]
    \centering
    \subfigure[Superellipse model.]{\label{Fig:ellipse}
    \includegraphics[width=0.75\linewidth]{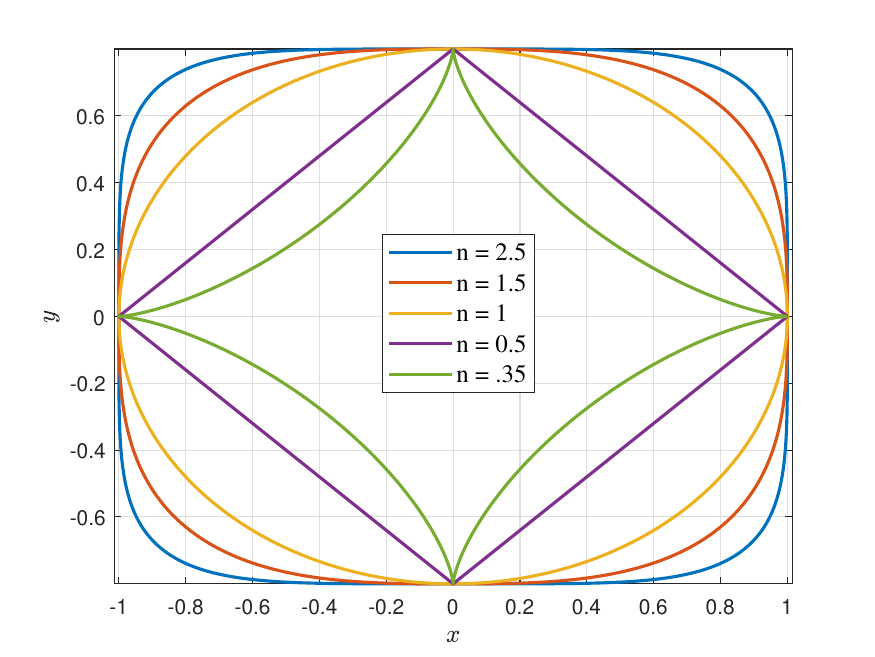}}
    \subfigure[B-spline parameterization model.]{\label{Fig:Bspline}
    \includegraphics[width=0.75\linewidth]{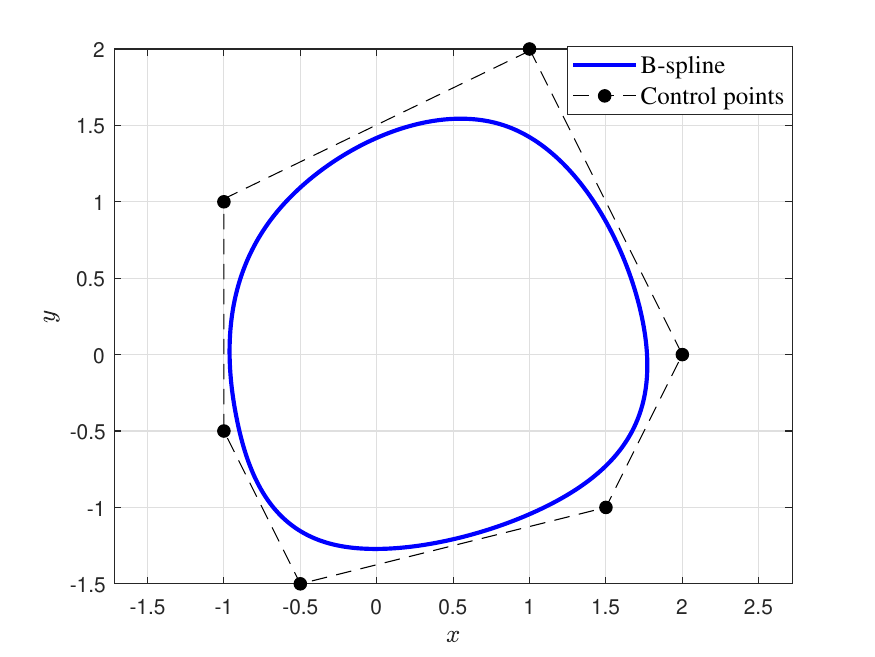}}
    \caption{\textcolor{black}{Superellipse and B-spline shape parameterizations.}}\label{Fig:para}
\end{figure}

To represent the shape of an arbitrary 2D closed geometry, a 
cubic B-spline parameterization can be used. Given $N_B$ control points 
$\{\bm{P}_1,\bm{P}_2,\dots,\bm{P}_{N_B}\}$, where each $\bm{P}_i$ is a 
2D vector $(x_i,y_i)^T$, a cubic B-spline $\boldsymbol\Gamma(\xi)$ of length $N_B$ in the 
domain $\xi\in[0,N_B]$ can be written as:
\begin{equation}
\label{eq:bspline_curve}
\boldsymbol\Gamma(\xi) \;=\; \sum_{i=1}^{N_B+3} \bm{\widetilde{P}}_i\, B_i(\xi),
\end{equation}
where $B_i(\xi)$ are the standard B-spline basis functions. To form a closed loop, 
we replicate the first three control points at the end to create a smooth curve that connects the end to 
the beginning. \textcolor{black}{A periodic cubic B-spline with six distinct control points, for example, yields a smooth closed contour, as shown in Fig.~\ref{Fig:Bspline}.} Increasing the number of control points $N_B$ enables the representation of more intricate geometries with finer local detail, at the expense of increased complexity and stricter identifiability constraints. Both types of curves in Fig.~\ref{Fig:para} are preferable to a pixel map when the object is well described by a smooth low-dimensional contour.

\subsection{Polygonal Shape Parameterization}
Another simple approach uses polygonal models parameterized by the $V$ locations of the polygon's vertices. This model requires $2V$ parameters corresponding to the $x$ and $y$ vertex coordinates. \textcolor{black}{For polygonal and spline models, the contour complexity ($V$ or $N_B$) is a model-order parameter. Below we show how to determine the polygon order through an MDL-type penalized concentrated likelihood.}

\subsection{ML Estimation of Source and Blockage Parameters}
\label{subsec:ml-framework}
Let $\bm{\gamma} \triangleq \begin{bmatrix}\bm{\theta}^{T} & \bm{\mu}^{T}\end{bmatrix}^{T}$
denote the vector of parameters to be estimated. For snapshot $t$, the received vector is
\begin{equation}
    \mathbf{s}_t
    =
    \mathbf{A}(\bm{\gamma})\mathbf{x}_t+
    \mathbf{n}_t,
    \qquad
    \mathbf{n}_t\sim\mathcal{CN}(\mathbf{0},\sigma^2\mathbf{I}_M),
    \label{eq:ml_data_model_blue}
\end{equation}
where $\sigma^2$ denotes the noise variance.
Define the set $\mathcal C_\gamma$ to contain feasible values for the parameters, including physical constraints such as positive ranges, positive object dimensions, admissible aspect ratios and
orientations, non-self-intersecting polygonal or spline contours, etc. 
\textcolor{black}{The size of the parameter vector is assumed known for the ML formulation, and the forward map must be locally injective after allowing for source-label permutation and an arbitrary complex scaling of each column of $\mathbf A$. Because the source covariance is unknown, a change in the scene that multiplies $\mathbf g_k(\boldsymbol\mu)$ by a constant across the array, such as a uniform attenuation or phase shift of source $k$, is absorbed into the source amplitude and carries no geometric information; all information about $\boldsymbol\mu$ comes from the variation of $g_{m,k}$ across the array elements. A physically unambiguous array manifold, sufficient spatial diversity, and excitation of distinct source subspaces are necessary conditions, but they are not sufficient for joint geometric identifiability. We can diagnose local identifiability from the positive definiteness and conditioning of the Fisher information matrix, but determining practical sufficient conditions for arbitrary objects, source distributions, and multiple blockages is difficult and remains an open problem.} For brevity, in what follows we will often write $\mathbf{A}(\bm{\gamma})$ as simply $\mathbf{A}$.

\subsubsection{Stochastic, or Unconditional, ML Model}
The stochastic ML model treats the source symbols as zero-mean {i.i.d.} random vectors with covariance $\mathbf{R}_x\succeq\mathbf{0}$. Then
\begin{align}
    \mathbf{s}_t &\sim\mathcal{CN}(\mathbf{0},\mathbf{R}_s) \\
    \mathbf{R}_s(\bm{\gamma},\mathbf{R}_x,\sigma^2)
    &=
    \mathbf{A}(\bm{\gamma})\mathbf{R}_x
    \mathbf{A}^{H}(\bm{\gamma})+
    \sigma^2\mathbf{I}_M.
\end{align}
For $N$ independent snapshots, the likelihood is
\begin{equation}\label{L_sto}
    L_{\rm s}(\bm{\gamma},\mathbf{R}_x,\sigma^2)
    =
    \frac{1}{\pi^{MN}\det(\mathbf{R}_s)^N}
    \exp\left(
        -\sum_{t=1}^N
        \mathbf{s}_t^{H}\mathbf{R}_s^{-1}\mathbf{s}_t
    \right).
\end{equation}
Equivalently, after defining the sample covariance
\begin{equation}
    \widehat{\mathbf{R}}_s
    =
    \frac{1}{N}\sum_{t=1}^N \mathbf{s}_t\mathbf{s}_t^{H},
\end{equation}
the stochastic model results in the objective
\begin{equation}
    J_{\rm s}(\bm{\gamma},\mathbf{R}_x,\sigma^2)
    =
    N\log\det\mathbf{R}_s
    +
    N\operatorname{tr}
    \left\{
        \mathbf{R}_s^{-1}\widehat{\mathbf{R}}_s
    \right\}, 
    \label{eq:stoch_nll_blue}
\end{equation}
and the stochastic ML estimate is therefore obtained from the constrained
optimization problem
\begin{equation}
    (\widehat{\bm{\gamma}}_{\rm s},
     \widehat{\mathbf{R}}_x,
     \widehat{\sigma}^2)
    =
    \mathop{\mathrm{arg\,min}}_{\substack{\bm{\gamma}\in\mathcal C_\gamma,\\
    \mathbf{R}_x\succeq\mathbf{0},\;\sigma^2>0}}
    J_{\rm s}(\bm{\gamma},\mathbf{R}_x,\sigma^2).
    \label{eq:stoch_mle_blue}
\end{equation}
\textcolor{black}{For fixed $\bm\gamma$ and full-column-rank $\mathbf A$ with $K<M$, the positive-semidefinite source covariance and noise power can be estimated in closed form. Let the columns of $\mathbf Q\in\mathbb C^{M\times K}$ form an orthonormal basis of $\operatorname{col}(\mathbf A)$ and let $\lambda_1\ge\cdots\ge\lambda_K$ be the eigenvalues of $\mathbf Q^H\widehat{\mathbf R}_s\mathbf Q$. For signal-subspace dimension $q\in\{0,\ldots,K\}$, set}
\begin{equation}\label{eq:profile_noise_new}
\textcolor{black}{\widehat\sigma_q^2=\frac{\operatorname{tr}(\widehat{\mathbf R}_s)
-\sum_{i=1}^{q}\lambda_i}{M-q},\quad
\lambda_q\ge\widehat\sigma_q^2\ge\lambda_{q+1},}
\end{equation}
\textcolor{black}{where $q$ is the largest index for which $\lambda_q>\widehat\sigma_q^2$ ($q=0$ if there is none). The resulting criterion is}
\begin{equation}\label{eq:profile_cost_new}
\textcolor{black}{J_{\rm s}^{\rm prof}(\bm\gamma)=N\left[
\sum_{i=1}^{q}\log\lambda_i+(M-q)\log\widehat\sigma_q^2+M\right].}
\end{equation}
\textcolor{black}{The estimated source covariance is positive semidefinite by construction. No full-rank assumption on $\mathbf R_x$ is needed, so correlated and coherent sources are handled by the same criterion, with $q<K$ when the estimated signal covariance is rank deficient. The simulations below use independent complex Gaussian source snapshots with unknown covariance and noise, unless stated otherwise.}

\subsubsection{Numerical Optimization}
The ML approach leads to a nonconvex optimization
problem since $\mathbf A(\boldsymbol\gamma)$ depends nonlinearly on $\bm{\gamma}$. Projected gradient or quasi-Newton iterations can find local minima via
\begin{equation}
    \bm{\gamma}^{(\ell+1)}
    =
    \Pi_{\mathcal C_\gamma}
    \left[
        \bm{\gamma}^{(\ell)}
        -
        \alpha_\ell
        \nabla_{\bm{\gamma}}J(\bm{\gamma}^{(\ell)})
    \right],
    \label{eq:correct_gradient_sign_blue}
\end{equation}
where \textcolor{black}{$J=J_{\rm s}^{\rm prof}$ is the concentrated ML criterion in \eqref{eq:profile_cost_new},} $\alpha_\ell>0$ is found via a line search, and $\Pi_{\mathcal C_\gamma}$ projects onto the feasible parameter set. If analytical derivatives
of the response are unavailable, the gradients can be approximated using central differences. The iterations
are stopped when the relative decrease in $J$ falls below a tolerance or when a prescribed iteration limit is reached.
\textcolor{black}{With $F_c$ frequencies, $K$ sources, $M$ receive elements, and $B$ non-overlapping blockages with $N_q$ integration points each, a direct forward evaluation requires $\mathcal O(F_cMKBN_q)$ kernel calculations, prior to the covariance factorization and nonlinear search. For $p$ unknown parameters, where $p$ grows linearly with the number of sources and the number of object vertices or control points, central numerical differences require approximately $2p$ extra forward evaluations per gradient, although analytical derivatives can reduce this cost. A fully digital array is assumed in the examples, but with $N_{\rm RF}<M$ RF chains the same likelihood can be applied to $\widetilde{\mathbf W}^H\mathbf s(t)$ and the effective response $\widetilde{\mathbf W}^H\mathbf A$, where $\widetilde{\mathbf W}=\mathbf W(\mathbf W^H\mathbf W)^{-1/2}$ whitens the compressed noise, provided $\mathbf W \in {\mathbb C}^{M \times N_{\rm RF}}$ is known and has full rank. The corresponding reduced-dimension FIM must then be evaluated, since such RF compression can compromise identifiability and diminish accuracy.}

\subsection{\textcolor{black}{MDL-Type Source and Shape Model-Order Selection}}\label{subsec:mdl_new}
\textcolor{black}{When the number of sources $K$ and the number of parameters in the object description $\boldsymbol\mu$ are unknown a priori, the optimization must be preceded by a model-order selection step. Here we describe an approach for this problem based on the well-known MDL principle \cite{wax1985detection} for the special case of the polygonal shape parameterization.
For $M$ array elements and $N$ independent snapshots, we first determine $K$ from the eigenvalues $\widehat\lambda_1\ge\cdots\ge\widehat\lambda_M$ of the sample covariance using the eigenvalue MDL statistic}
\begin{equation}\label{eq:source_mdl_new}
\textcolor{black}{\operatorname{MDL}_{\rm src}(k)=
N(M-k)\log\frac{\overline\lambda_k}{\widetilde\lambda_k}
+\frac{k(2M-k)}{2}\log N,}
\end{equation}
\textcolor{black}{where $\overline\lambda_k=\sum_{i=k+1}^{M}\widehat\lambda_i/(M-k)$ and $\widetilde\lambda_k=(\prod_{i=k+1}^{M}\widehat\lambda_i)^{1/(M-k)}$ are the arithmetic and geometric means of the smallest $M-k$ eigenvalues, respectively.}
\textcolor{black}{Given $\widehat K$, the polygonal vertex count $V$ is chosen from a finite candidate set by minimizing}
\begin{equation}\label{eq:vertex_mdl_new}
\textcolor{black}{\operatorname{MDL}_{\rm poly}(V)=2\min_{\bm\gamma\in\mathcal C_V}
J_{\rm s}^{\rm prof}(\bm\gamma)+2V\log N} ,
\end{equation}
\textcolor{black}{where $2V$ represents the number of vertex coordinate parameters. The polygon search is conditional on
$\widehat K$; each candidate is evaluated by minimizing the same
concentrated negative log-likelihood before adding the
penalty in \eqref{eq:vertex_mdl_new}. The performance of the above model-order selection approach is investigated in Section~VI-F.}

\subsection{Shape Initialization}
\label{subsec:init-guess}
A good initial estimate for ${\boldsymbol{\mu}}$ significantly increases the likelihood of convergence to the global optimum. To discuss how to find such an initialization, let $U_{ij}(x',y')$ denote the diffraction field measured by the receive antenna located at $(x'_i,y'_j)$.
Taking a triangle blockage as an example, the task is to obtain an initial guess for the three vertices that define the object:
\(\boldsymbol{\mu}^{(0)}=[x_1,y_1,x_2,y_2,x_3,y_3]^T\). We first threshold the magnitude of the received data to extract a crude contour. Define the normalized receive magnitude
\begin{equation}
{M}_{ij}=\frac{\bigl|U_{ij}\bigl|-M_{\min}}{M_{\max}-M_{\min}},\end{equation}
where $M_{\min}=\min_{i,j}\bigl|U_{ij}\bigl|$ and $M_{\max}\!=\!\max_{i,j}\bigl|U_{ij}\bigl|$. Since the magnitude drops near the shadow edge, we consider the window  
\([\,\zeta_{\min},\zeta_{\max}]\subset(0,1)\) and identify the elements that correspond to the shadow edge using a binary mask \(E\):
\begin{equation}
E_{ij}=\begin{cases}
1,& \zeta_{\min}\le{M}_{ij}\le\zeta_{\max},\\
0,& \text{otherwise}.
\end{cases}
\end{equation}
Second, using the 8-connected border-tracing algorithm~\cite{gonzalez2004digital}, we find the largest connected component in $E$ and extract its outer boundary. We can then select suitable initial vertices by fitting the polygon to the boundary points. 

Fig.~5 shows an example of the above procedure for a $20 \times 20$ UPA receive array with a triangular blockage.
We choose suitable threshold values and then extract a closed
contour illustrated by the solid pink line. After selecting three
vertices, the initial guess is obtained as the black triangle.
Fig.~5 illustrates that non-RF-informed image processing can only provide a crude estimate of the blockage shape without
exploiting information about the diffraction field.

\begin{figure}[t!]
\label{Figinit}
\begin{center}
\includegraphics[scale=0.45]{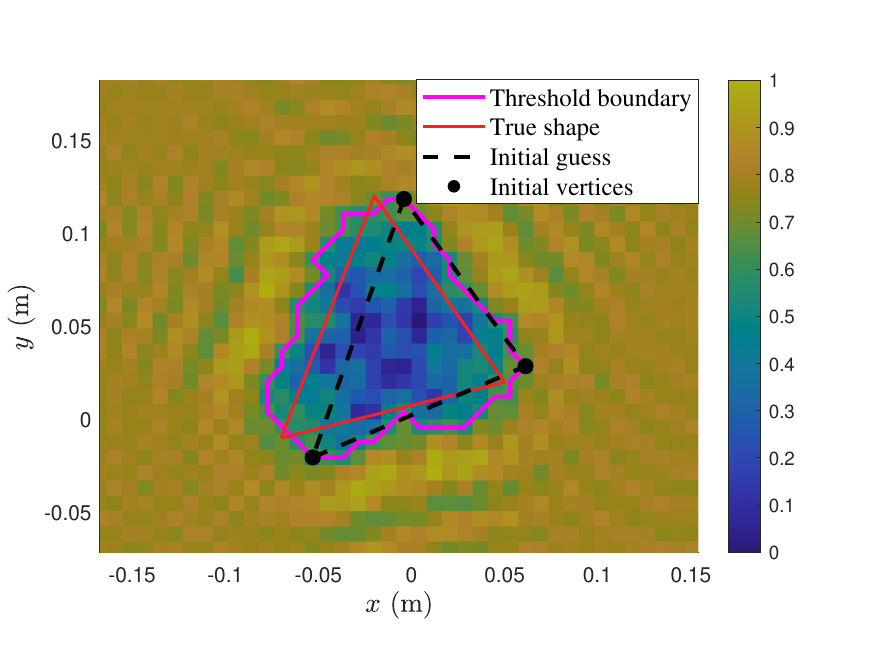}
\caption{Illustration of the coarse shape-initialization procedure.}
\end{center}
\end{figure}

\section{Cram\'er--Rao Bound Analysis}
\label{sec:CRB}
To characterize the fundamental limits for estimating 
$\bm{\gamma}=[\bm{\theta}^{T}\;\bm{\mu}^{T}]^{T}$, we use the CRB corresponding to the stochastic ML model \cite{17564,60109,120802,kay1993fundamentals}. The bounds corresponding to the DoA and shape parameters can be found from 
\begin{align}
    \operatorname{CRB}(\bm{\theta})
    &=
    \left(
        \mathbf{F}_{\theta\theta}
        -
        \mathbf{F}_{\theta\mu}
        \mathbf{F}_{\mu\mu}^{-1}
        \mathbf{F}_{\mu\theta}
    \right)^{-1}
    \nonumber\\
    \operatorname{CRB}(\bm{\mu})
    &=
    \left(
        \mathbf{F}_{\mu\mu}
        -
        \mathbf{F}_{\mu\theta}
        \mathbf{F}_{\theta\theta}^{-1}
        \mathbf{F}_{\theta\mu}
    \right)^{-1},
    \label{eq:crb_schur_blue}
\end{align}
which are Schur complements of the partitioned Fisher Information Matrix (FIM) \cite{kay1993fundamentals}:
\begin{equation}
    \mathbf{F}_{\gamma\gamma}
    =
    \begin{bmatrix}
        \mathbf{F}_{\theta\theta} & \mathbf{F}_{\theta\mu}\\
        \mathbf{F}_{\mu\theta} & \mathbf{F}_{\mu\mu}
    \end{bmatrix}.
    \label{eq:partitioned_fim_blue}
\end{equation}

\subsection{Stochastic CRB}

\iffalse
For the deterministic ML model, for each snapshot we define the Jacobian
\begin{equation}
    \mathbf{B}_t
    =
    \left[
        \frac{\partial \mathbf{A}}{\partial \gamma_1}\mathbf{x}_t,
        \ldots,
        \frac{\partial \mathbf{A}}{\partial \gamma_{P_\gamma}}\mathbf{x}_t
    \right]
    \in\mathbb{C}^{M\times P_\gamma},
    \label{eq:Bt_det_blue}
\end{equation}
where $P_\gamma$ is the total number of parameters. The FIM for
$\bm{\gamma}$ is given by the standard
conditional-model result \cite{60109}
\begin{equation}
    \mathbf{F}_{\gamma\gamma}^{\rm d}
    =
    \frac{2}{\sigma^2}
    \operatorname{Re}
    \left\{
        \sum_{t=1}^N
        \mathbf{B}_t^{H}
        \mathbf{P}_{\mathbf{A}}^{\perp}
        \mathbf{B}_t
    \right\},
    \label{eq:det_fim_blue}
\end{equation}
with $\mathbf{P}_{\mathbf{A}}^{\perp}$ as in
\eqref{eq:det_concentrated_cost_blue}. Only the components of the
parameter derivatives that lie outside the column space of
$\mathbf{A}$ carry information about $\bm{\gamma}$.
\fi

For the stochastic ML model, the FIM for known $\mathbf{R}_x$ and $\sigma^2$ is given by the
Slepian--Bangs formula for proper complex Gaussian data
\cite{kay1993fundamentals,stoica2005spectral}:
\begin{equation}
    [\mathbf{F}_{\gamma\gamma}^{\rm s}]_{ij}
    =
    N\operatorname{tr}
    \left\{
        \mathbf{R}_s^{-1}
        \dot{\mathbf{R}}_i
        \mathbf{R}_s^{-1}
        \dot{\mathbf{R}}_j
    \right\},
    \qquad
    \dot{\mathbf{R}}_i
    \triangleq
    \frac{\partial\mathbf{R}_s}{\partial\gamma_i}.
    \label{eq:stoch_fim_element_blue}
\end{equation}
The CRBs for $\bm{\theta}$
and $\bm{\mu}$ follow by partitioning the FIM as in
\eqref{eq:partitioned_fim_blue} and applying
\eqref{eq:crb_schur_blue}. \textcolor{black}{When $\mathbf R_x$ and $\sigma^2$ are estimated from the same data, their influence is removed from the FIM through the nuisance-parameter Schur complement below.}
The only model-specific components in
\eqref{eq:stoch_fim_element_blue} are the
derivatives $\partial\mathbf{A}/\partial\gamma_i$. For a rectangular blockage, these are available in
closed form by differentiating \eqref{a},
\eqref{eq:g_rect_compact}, and \eqref{eq:g_rect_expanded}; for
B-spline, superellipse, and polygonal parameterizations, they are
computed by numerical differentiation of the finite-domain Fresnel
integral \eqref{eq:babinet_fresnel_scaled}.

{\color{black}
When $\mathbf{R}_x$ and $\sigma^2$ are unknown, they are handled as
additional nuisance parameters rather than replaced by unconstrained
estimates that may violate positive semidefiniteness. Let
$\bm{\beta}=[\operatorname{vech}_{\mathbb R}(\mathbf{R}_x)^{T}\;\;\sigma^2]^{T}$
collect a nonredundant real parameterization of the Hermitian source
covariance (its real diagonal entries and the real and imaginary parts
of its strictly lower-triangular entries) together with the noise
variance. The full FIM over $[\bm{\gamma}^{T}\;\bm{\beta}^{T}]^{T}$ has entries given by
\eqref{eq:stoch_fim_element_blue}, with
$\partial\mathbf{R}_s/\partial\beta_i
=\mathbf{A}\,(\partial\mathbf{R}_x/\partial\beta_i)\,\mathbf{A}^{H}$
for the covariance components and
$\partial\mathbf{R}_s/\partial\sigma^2=\mathbf{I}_M$. The equivalent
FIM for the structural parameters is then the Schur complement
\begin{equation}
    \mathbf{F}_{\gamma|\beta}
    =
    \mathbf{F}_{\gamma\gamma}
    -
    \mathbf{F}_{\gamma\beta}
    \mathbf{F}_{\beta\beta}^{\dagger}
    \mathbf{F}_{\beta\gamma},
    \label{eq:nuisance_schur_blue}
\end{equation}
where $\dagger$ denotes the Moore--Penrose inverse, and the CRBs for $\bm{\theta}$ and $\bm{\mu}$ follow from
\eqref{eq:partitioned_fim_blue} and \eqref{eq:crb_schur_blue} as
before.}
\subsection{Identifiability of Object Scale and Range}
\label{subsec:identifiability}
\textcolor{black}{A full-rank $\mathbf{F}_{\gamma|\beta}$ in~\eqref{eq:nuisance_schur_blue} establishes only local rather than global identifiability. We illustrate one geometric near-ambiguity below, but general sufficient conditions for complex object shapes and unknown source waveforms are challenging and remain as future work.}

The Fresnel/Babinet representation \eqref{eq:babinet_fresnel_scaled}
reveals a near-ambiguity between the transverse 
scale of the object and its range. Suppose the blockage parameters are decomposed into an
overall transverse scale $s$ (e.g., the width $d$ of a rectangle\textcolor{black}{, together with any offset of its center}) and
scale-free shape parameters, and consider the \textcolor{black}{transformation}
\begin{equation}
    s\mapsto \alpha s,
    \qquad
    R\mapsto \alpha^2 R,
    \qquad
    \alpha>0,
    \label{eq:scale_range_family}
\end{equation}
which leaves the Fresnel number $F=d^2/(\lambda R)$, and hence the
scaled support $D_s(\bm\mu,R)$, unchanged. Under
\eqref{eq:scale_range_family}, the shifted receive coordinate of a broadside source scales as $u_{m,k}'\mapsto u_{m,k}'/\alpha$, and the identical Fresnel pattern is merely dilated by the factor $\alpha$ across the fixed physical aperture. Unambiguously resolving $s$ and $R$ is therefore achieved only by (i) this dilation of the pattern relative to the fixed array geometry, and (ii) the DoA-dependent lateral offsets $Rq_{x,k}$ and $Rq_{y,k}$, which transform differently under \eqref{eq:scale_range_family} for oblique sources. When the aperture is small in the dimensionless coordinate $u=\sqrt{2/(\lambda R)}\,x$ and hence captures only a few Fresnel oscillations, the columns of the Jacobian $\partial\mathbf{a}/\partial s$ and $\partial\mathbf{a}/\partial R$ become nearly collinear, the Schur complements in \eqref{eq:crb_schur_blue} become ill-conditioned, and
the CRBs for scale and range can become large even at high SNR as the diffraction features flatten across the aperture. The conditioning improves with electrically larger apertures, with multiple sources at distinct DoAs, and \textcolor{black}{at shorter wavelengths, for which the same physical aperture spans more Fresnel fringes. Note that \eqref{eq:scale_range_family} leaves $F$ unchanged at every wavelength, so wideband measurements help through this larger dimensionless aperture and not by removing the invariance. The ambiguity would be exact only if the receive aperture were scaled by $\alpha$ together with the geometry, which is the constant-Fresnel-number scaling \eqref{eq:const_F_map} with $\lambda'=\lambda$ and $R'=\alpha^2R$.} For a fixed physical aperture, it manifests itself as ill-conditioning rather than exact nonidentifiability.

\section{Simulation Results and Discussion}
\label{sec:sim}
In this section, we provide simulation results to validate the proposed diffraction model and evaluate the ML estimation performance for the blockage parameters.

\subsection{Simulation Setup}
\textcolor{black}{We consider several complementary scenarios: far-field illumination of thin polygonal screens, known far-field illumination of a dielectric superellipse with unknown thickness/material and tilt, and near-field illumination of a finite-thickness rectangular object. Receive arrays are uniform linear or rectangular planar arrays. Source signals are proper complex Gaussian unless explicitly specified, while the stochastic ML method treats the source covariance matrix and receiver-noise power as unknown nuisance parameters.}

The studies below use carrier frequencies between \textcolor{black}{$5.8$} and $60$~GHz. Since the paraxial diffraction pattern depends on the scene geometry only through the Fresnel number $F=d^2/(\lambda R)$ and the dimensionless aperture coordinates (Section~\ref{sec:sim}-\ref{subsec:scaling}), each configuration below is representative of an entire equivalence class of scenarios at other carrier frequencies. For example, the $60$~GHz shape-estimation scenario of Section~\ref{sec:sim}-\ref{subsec:shape_est} (a $\sim 10$~cm object at $R=0.4$~m) produces the same diffraction pattern \textcolor{black}{on the same physical aperture as a $10$~cm object at approximately $0.67$~m from the array at $100$~GHz, since $\lambda R$ is the same in both cases.} The lower carrier frequencies used here keep the full-wave validation computationally tractable.

\subsection{Validation of the Diffraction Model}

\begin{figure}[!t]
    \centering
    \subfigure[HFSS setup.]{\label{Fig:HFSS1}
    \includegraphics[width=0.6\linewidth]{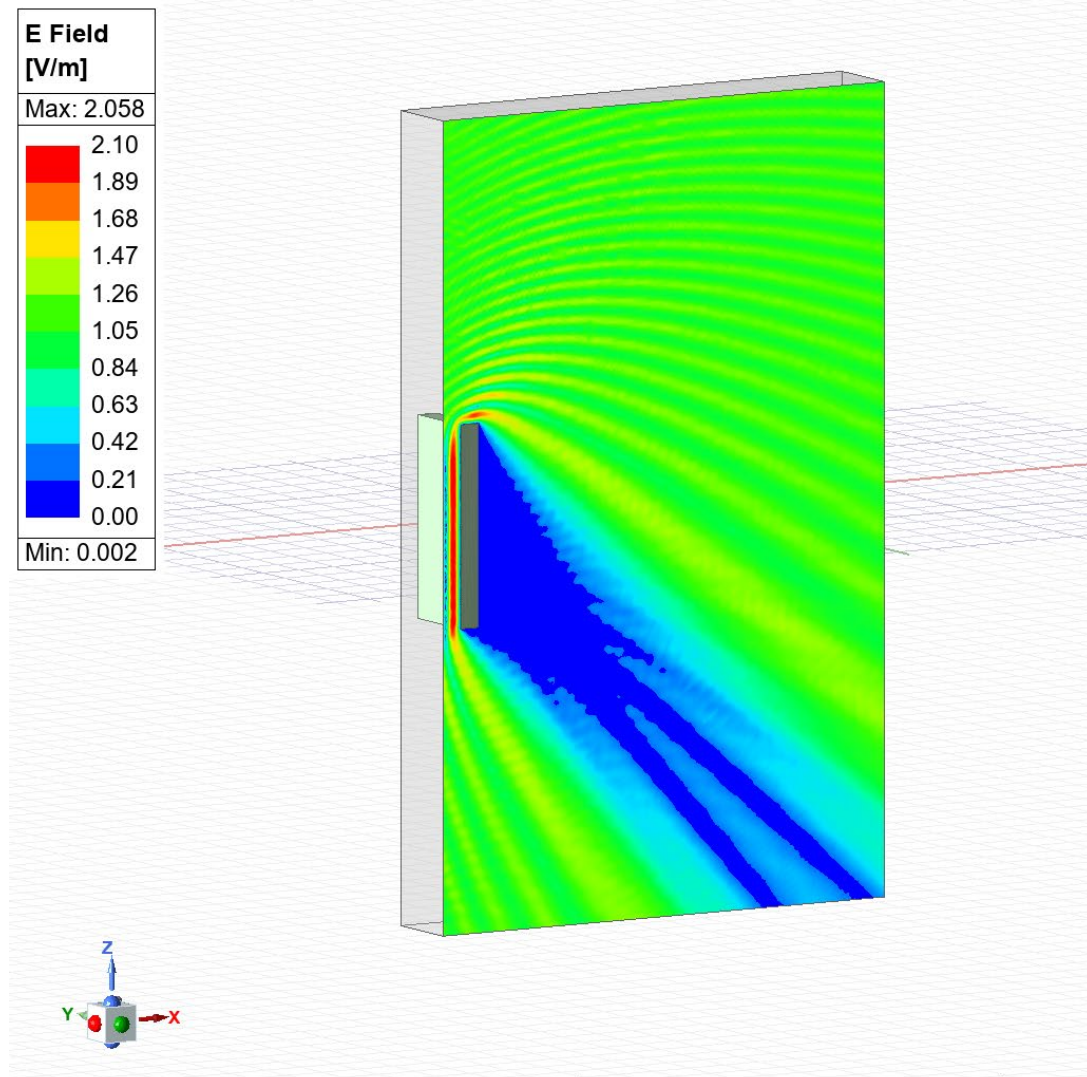}} 
    \subfigure[Analytical model vs. HFSS.]{\label{Fig:HFSS2}
    \includegraphics[width=0.7\linewidth]{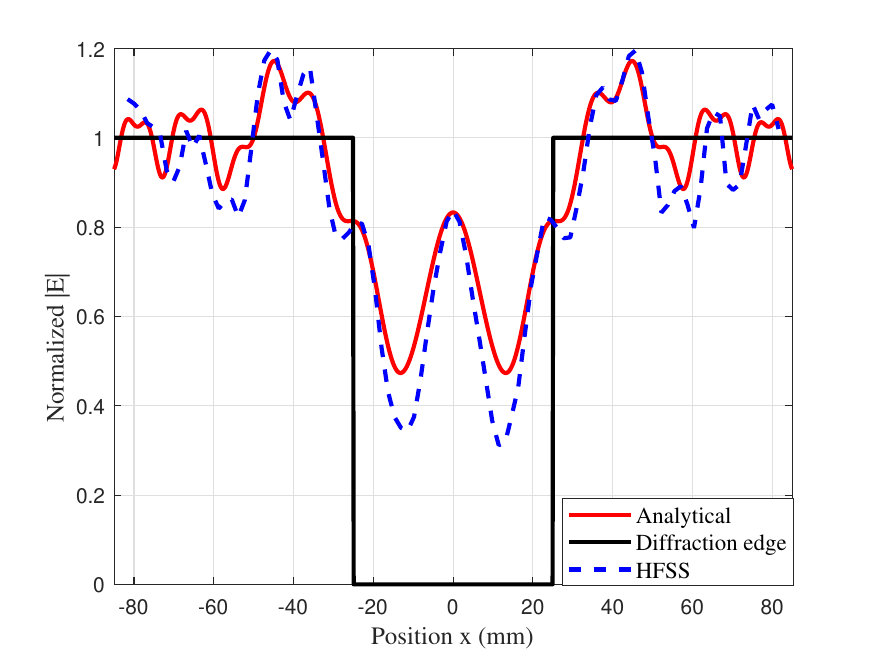}}
    \caption{Validation of the proposed analytical model using HFSS.}\label{Fig:HFSS}
\end{figure}

To verify the diffraction model, we compare the analytical received field to a full-wave simulation via finite element methods. Fig.~\ref{Fig:HFSS} illustrates an Ansys high-frequency structure simulator (HFSS) setup for full-wave simulation of signal diffraction from a rectangular prism. The thickness of the prism is $5$ mm, which is negligible compared to its other dimensions. \textcolor{black}{In Fig.~\ref{Fig:HFSS1}, a 50 GHz plane wave impinges on the object with a $45^\circ$ DoA, while the numerical comparison in Fig.~\ref{Fig:HFSS2} assumes normal incidence. The accompanying one-dimensional analytical calculation uses a strip width of $0.05$~m and an observation distance of $0.11$~m.} As the signal is partially blocked by the perfectly conducting object, a diffraction pattern is created and the magnitude of the electric field fluctuates near the shadow edges.
In Fig.~\ref{Fig:HFSS2}, we see that the analytical model given in~\eqref{a} closely matches the simulation results. Due to the finite boundary limitation in HFSS, some discrepancies occur around the extreme points of the electric field magnitude, but the mean amplitude difference across the array is less than about 1~dB.

\subsection{Fresnel Number and Scaling Laws}\label{subsec:scaling}
The physics-based diffraction channel model provides a powerful tool to understand how diffraction patterns scale with object size and range. Consider diffraction from a blockage with dimension $d$, e.g., a rectangular strip of width $d$, located at $z = 0$ whose diffraction pattern is observed by a ULA at $z = R$. In the following, we discuss the scaling behavior of the signal diffraction pattern under either a global uniform scaling or a constant-Fresnel number scaling.

\subsubsection{Uniform Scaling}

Based on the model in \eqref{eq:diff_op}, the array response at a point $(x',R)$ on the linear array can be written as:
\begin{equation}
        \textcolor{black}{a(x') \;\propto\; \int_{0}^{d}
    \frac{R}{\rho^{2}}\,e^{\,j\,\frac{2\pi}{\lambda}\,\rho}
    \,\mathrm{d}x,\qquad \rho=\sqrt{(x - x')^{2} + R^{2}}.}
\end{equation}
A global scaling of all dimensions by the factor $\alpha$, i.e.,
\[
    d \,\mapsto\, \alpha\,d, \quad
    R \,\mapsto\, \alpha\,R, \quad
    x' \,\mapsto\, \alpha\,x', \quad
    \lambda \,\mapsto\, \alpha\,\lambda,
\]
leaves the diffraction integral invariant.
Thus, the shape of the diffraction pattern is preserved up to a
stretch in the transverse coordinate by the factor $\alpha$.

\subsubsection{Constant-Fresnel Number Scaling}

As long as a given blockage is within the Fresnel region of the receive array, keeping \textcolor{black}{$F=d^2/(\lambda R)$} constant ensures that the observed diffraction pattern remains the same. This is because, \textcolor{black}{by \eqref{eq:babinet_fresnel_scaled}, within the paraxial region the array response depends on the geometry only through $F$ and the dimensionless coordinate $u$.} Preserving $F$ ensures the same Fresnel diffraction pattern shape in $u$-space. Depending on how $\lambda$ and $R$ scale, the real-space pattern may {stretch} or {compress}.  In particular, consider the two settings $(d,\lambda,R)$ and $(d',\lambda',R')$. The coordinate mapping that keeps the diffraction pattern constant is:
\begin{equation}\label{eq:const_F_map}
   x' \;=\; x \;\sqrt{\frac{\lambda' R'}{\lambda R}}.
\end{equation}
Thus, (i) if $\lambda' R' > \lambda R$, the pattern in $x'$-space is enlarged, (ii) if $\lambda' R' = \lambda R$, the pattern is the same size in $x$, and (iii) if $\lambda' R' < \lambda R$, the pattern is compressed.

\textcolor{black}{\emph{Design implications:} For a transverse object size $d$ at range $R$, the Fresnel number $F=d^2/(\lambda R)$ identifies the region in which edge fringes provide informative variations across the aperture. Changing operating frequency or range shifts these fringes, and a receive aperture too small to sample their spatial variation yields poorly conditioned geometric derivatives. Reducing the element spacing and increasing the aperture width can resolve such issues, multiple incidence angles and shorter wavelengths can help distinguish object scale from range, and use of multiple separated frequencies can help distinguish thickness from dielectric contrast.}

\begin{figure}[t!]
    \begin{center}
        \includegraphics[scale=0.45]{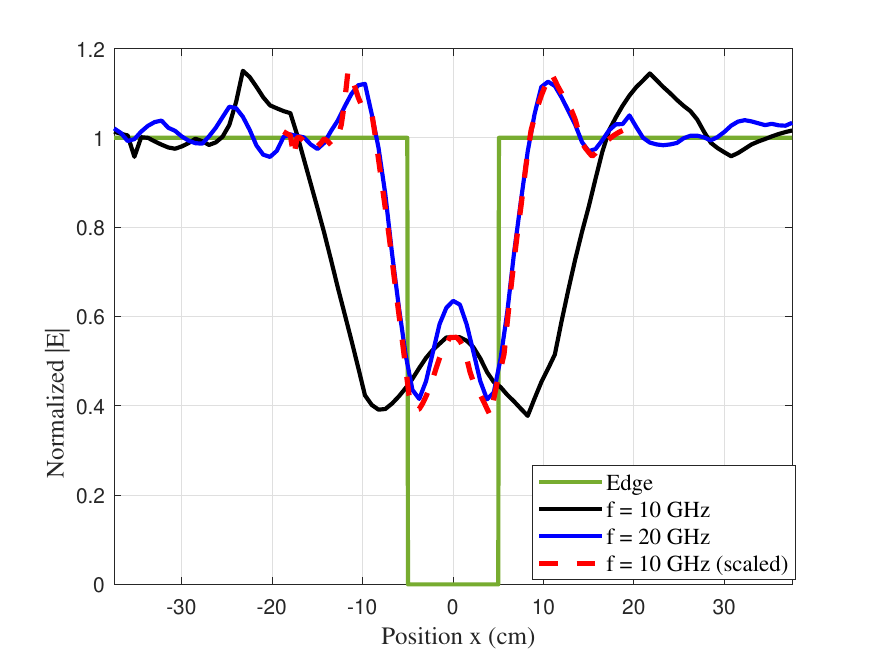}
        \caption{\textcolor{black}{Uniform scaling law illustrated via HFSS.}}
        \label{Fig:scale}
    \end{center}
\end{figure}

\subsubsection{HFSS Validation}
In Fig.~\ref{Fig:scale}, we use a full-wave simulation to \textcolor{black}{illustrate the uniform scaling law} for a ULA with a rectangular blockage of width $d$. The black curve is the diffraction pattern for $d = 20$ cm, $\lambda = 3$ cm, and $R = 0.5$ m. The blue curve corresponds to a uniform scaling of $\alpha=0.5$, where $d' = 10$ cm, $\lambda' = 1.5$ cm, and $R' = 0.25$ m. To see that the blue response is a compressed version of the one in black, we appropriately scale the black response and plot it as a dashed red line, which closely overlaps the blue response. \textcolor{black}{The constant-Fresnel-number law follows analytically from \eqref{eq:babinet_fresnel_scaled}.} Beyond providing physical insight, these scaling laws allow the estimation results reported in this section to be mapped directly to mmWave/THz configurations with the same Fresnel number, and they underlie the scale-range identifiability discussion in Section~\ref{sec:CRB}-\ref{subsec:identifiability}.

\begin{figure}[!t]
    \centering
    \subfigure[Received magnitude.]{\label{Fig:spline_pattern}
    \includegraphics[width=0.6\linewidth]{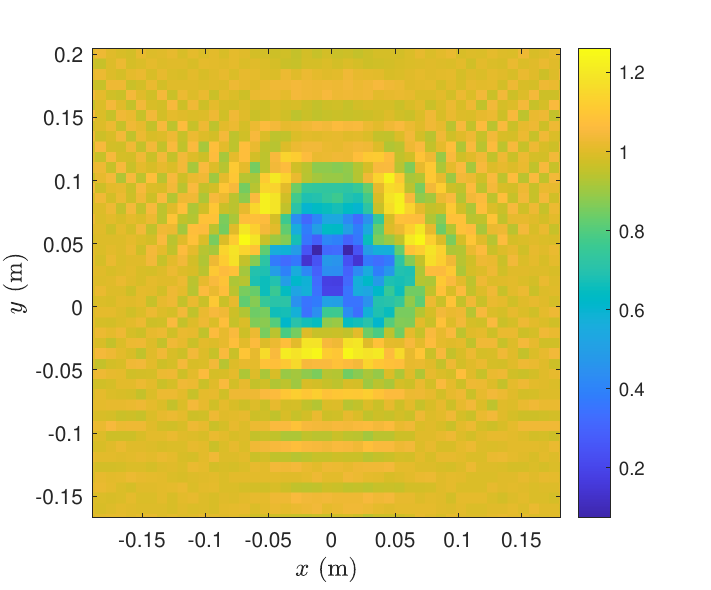}} 
    \subfigure[True and estimated shapes.]{\label{Fig:spline_shape}
    \includegraphics[width=0.6\linewidth]{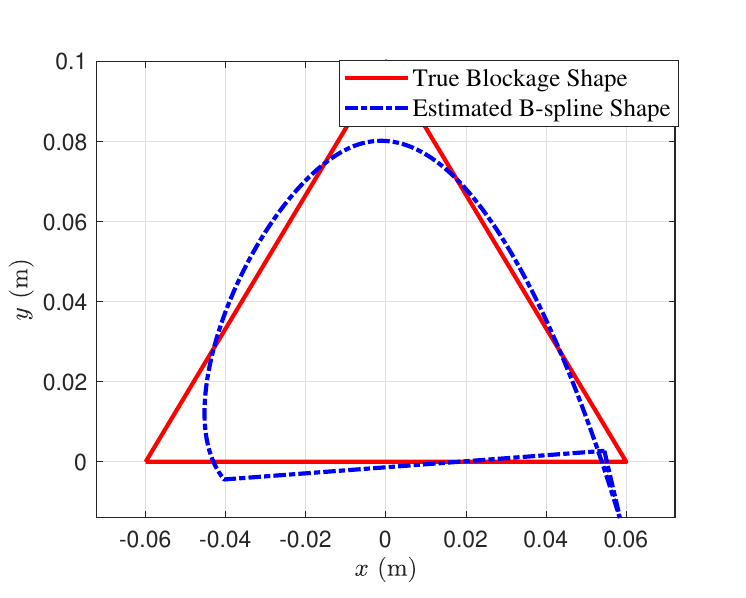}}
    \caption{Shape estimation using the B-spline parameterization model.}\label{Fig:spline}
\end{figure}

\subsection{Shape Estimation Example}\label{subsec:shape_est}
In Fig.~\ref{Fig:spline}, we use the B-spline model of Section~\ref{sec:Shape}-\ref{sub:bspline} to image a triangular blockage located $0.4$~m from a $50\times 50$ UPA. The source transmits at $60$~GHz with normal incidence on the UPA. In this case the source DoA and blockage range are known, and we estimate the shape using a B-spline with 4 control points. The ML estimator is initialized with four control points around a circle centered at $(0,0)$ with a diameter of $0.1$ m. The algorithm then evaluates $A(\theta = 0, \bm\mu)$ by numerical integration based on Babinet's principle. \textcolor{black}{The B-spline control points are fit by minimizing the complex-field least-squares error for a single noisy complex-field observation over the $50\times 50$ receive array with a single temporal snapshot. The incident field amplitude is $U_0=1$ and independent circular complex Gaussian noise is added with standard deviation $\sigma_0=0.01$, corresponding to a nominal unobstructed-field SNR of 40~dB.}
Fig.~\ref{Fig:spline_pattern} plots the magnitude of the received signal on a linear scale, illustrating the oscillatory variations due to diffraction.  Fig.~\ref{Fig:spline_shape} shows the true object shape in red, and the estimated B-spline after ML estimation. We observe that a reasonably accurate shape estimate is obtained with a small number of control points.

\begin{figure*}[!t]\centering
\includegraphics[width=0.8\textwidth]{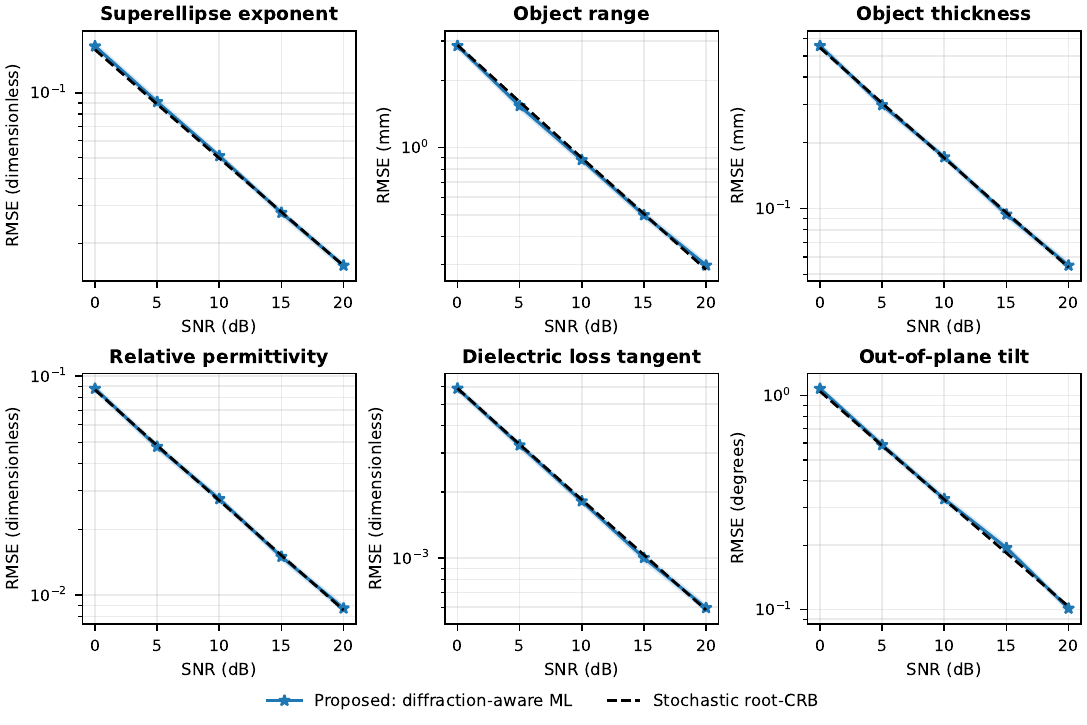}
\caption{\textcolor{black}{RMSE of superellipse exponent, object range, thickness, dielectric parameters, and out-of-plane tilt (1000 trials per SNR; 180 snapshots at each of three frequencies). Solid starred curves: ML estimates; dashed curves: stochastic root-CRBs; shading: 95\% confidence intervals on RMSE.}}
\label{fig:superellipse_primary_1000}
\end{figure*}

\subsection{\textcolor{black}{Joint Superellipse, Material, and Tilt Estimation}}\label{subsec:superellipse_new}
\textcolor{black}{To extend the thin opaque-screen model, we first consider a $13\times9$ rectangular array illuminated by three far-field sources at known directions, with independent measurements at $5.8$, $7.8$, and $10.4$~GHz. An inclined dielectric superellipse is present, described by 11 real unknowns: horizontal and vertical center coordinates, two semi-axes, in-plane rotation, continuous superellipse exponent $n$, range, axial thickness $h$, real relative permittivity $\varepsilon_r'$, loss tangent $\tan\delta$, and out-of-plane tilt about the vertical axis. The forward response uses the obliquity-aware scalar diffraction integral with a coherent local dielectric-slab transmission approximation. We estimate the 11 parameters jointly by stochastic ML with unknown source covariance and noise at each frequency.}

\textcolor{black}{ The material is described by a
nondispersive dielectric approximation rather than a categorical
material classifier. The dielectric loss tangent
$\tan\delta=\varepsilon_r''/\varepsilon_r'$ measures the
ratio of the loss-producing imaginary part of the effective
relative permittivity to its energy-storing real part.
The three frequencies provide different phase and
attenuation responses for the same object, allowing the
estimator to exploit frequency diversity and distinguish
a thicker, lower-contrast slab from a thinner, higher-contrast
slab. Strong dispersion or conducting edges 
requires a different physical parameterization.}

\textcolor{black}{Fig.~\ref{fig:superellipse_primary_1000} shows RMSEs from 1000 Monte Carlo trials per SNR and 180 snapshots per frequency. The shaded regions indicate uncertainty due to the finite trials (95\% confidence intervals). The RMSEs for the object center, semi-axes, and orientation are not included but show similar behavior. The FIM is full rank for the assumed multi-frequency probing, so the tilt is locally identifiable within the adopted scalar model. The proposed approach provides object parameter estimates that achieve the bound for all SNRs, illustrating its potential for accurate environmental sensing.}

\iffalse
\begin{figure*}[!t]\centering
\includegraphics[width=0.96\textwidth]{RevisionFigures/superellipse_shape_compacter.pdf}
\caption{\textcolor{black}{RMSE and root-CRB for the vertical center, minor semi-axis length, and in-plane orientation for the same experiment as Fig.~\ref{fig:superellipse_primary_1000}.}}
\label{fig:superellipse_shape_1000}
\end{figure*}

\textcolor{black}{The additional geometric error curves in
Fig.~\ref{fig:superellipse_shape_1000} are included to clarify which
aspects of the inferred shape are being tested.  The horizontal and
vertical center parameters determine the projected object's position,
the two semi-axes control its overall extent and aspect ratio, and
the in-plane orientation rotates the contour without tilting its
normal.  The continuous exponent $n$ controls the contour's
``squareness'' and is distinct from the discrete polygon vertex
count in the following example.  In contrast, out-of-plane tilt
changes both the projected outline and the path/incidence geometry.
The selected three-frequency illumination makes the equivalent FIM
full rank, but the strong thickness--permittivity covariance means
that the small individual RMSEs are a property of this observation
configuration and not a blanket material-identification claim.}
\fi

\begin{figure}[!t]\centering
\includegraphics[width=0.98\linewidth]{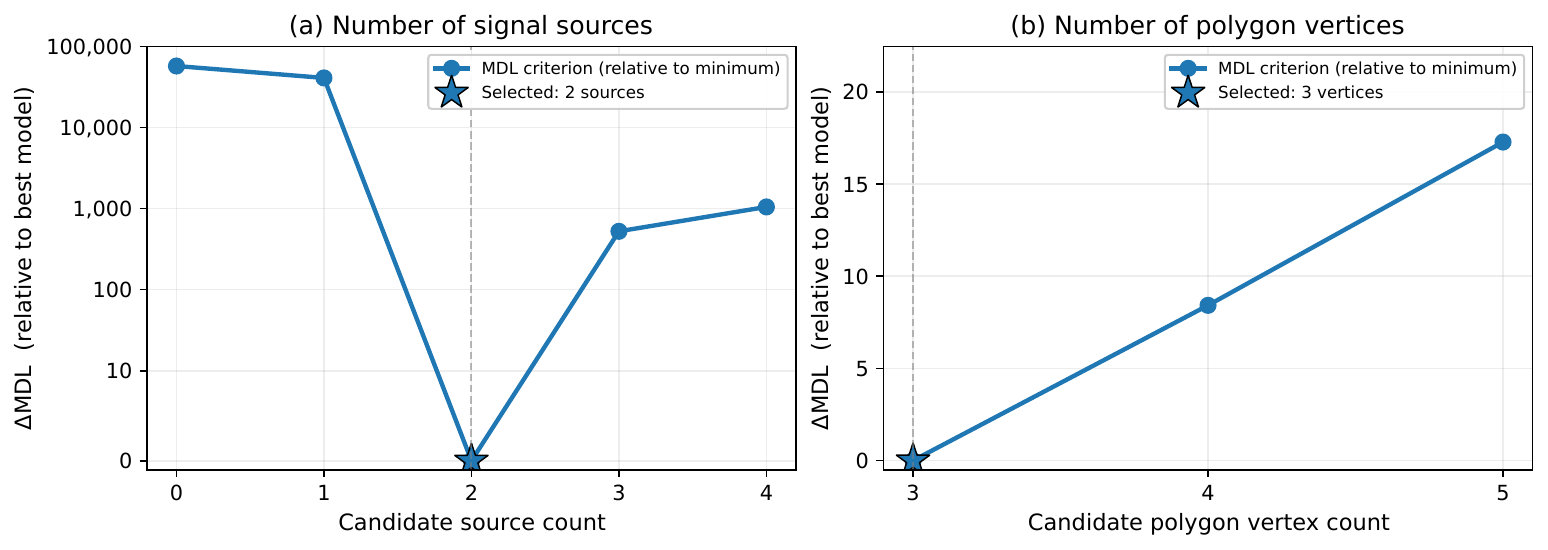}
\caption{\textcolor{black}{Representative trial illustrating the MDL-based model-order decisions for source count ($K=2$) and triangle vertex count ($V=3$). Both criteria are shifted to have minimum zero for display.}}
\label{fig:mdl_orders_new}
\end{figure}

\subsection{\textcolor{black}{Polygonal Object and Joint Model-Order Selection}}\label{subsec:polygon_new}
\textcolor{black}{Next we consider a triangular object using the polygonal model, estimating source DoAs, object range, and three 2D vertex locations. Two independent far-field sources at DoAs $-13^\circ$ and $17^\circ$ illuminate an opaque triangle with vertices $(-0.064,-0.045)$, $(0.066,-0.041)$, and $(0.002,0.083)$~m at range 0.46~m. The receive array is a $12\times12$ $\lambda/2$-spaced UPA at 5.9~GHz. The source and noise statistics are unknown and must be estimated. The MDL criteria in~\eqref{eq:source_mdl_new} and~\eqref{eq:vertex_mdl_new} are used to determine $K$ and $V$ before ML estimation, and are evaluated for $K\in\{0,1,2,3,4\}$ sources and polygon orders $V\in\{3,4,5\}$. The MDL approach correctly selected two sources in all 800 trials and three vertices in 794/800 trials (99.25\%). Fig.~\ref{fig:mdl_orders_new} displays a representative trial of MDL for an SNR of 10~dB, and Fig.~\ref{Fig:poly} shows the associated local likelihood convergence. Cases where the wrong order is selected are not included in the RMSE calculations.}

\iffalse
\textcolor{black}{For context, the original polygon demonstration in
Fig.~\ref{Fig:poly_legacy} used one source at known DoA, a 20$\times$20
planar array at 6~GHz, and a thin opaque triangle, and estimated
only the triangle vertices and object range.  The original
example initialized the range at 0.5~m and the contour using the
method illustrated in Fig.~\ref{Fig:init}. The iteration trace in
Fig.~\ref{Fig:poly_legacycost} illustrates the local reduction of
the original likelihood, and Fig.~\ref{Fig:poly_legacyshape} shows
the corresponding recovered outline.  These are the original
illustrations, not outputs of the newer MDL-selected experiment;
they are retained because they make the optimization procedure
tangible.  The earlier 60-trial, 20-snapshot RMSE study is
superseded by the more extensive experiment below.}

\begin{figure}[!t]
\centering
\subfigure[Original cost-function trace.]{\label{Fig:poly_legacycost}
\includegraphics[width=0.8\linewidth]{EPSs/Cost.pdf}}
\subfigure[Original true and estimated shapes.]{\label{Fig:poly_legacyshape}
\includegraphics[width=0.8\linewidth]{EPSs/Shape_est.pdf}}
\caption{\textcolor{black}{Original polygonal shape and range experiment with source direction assumed known; distinct from the subsequent MDL-selected joint-source experiment.}}
\label{Fig:poly_legacy}
\end{figure}
\fi

\begin{figure}[!t]\centering
\includegraphics[width=.8\linewidth]{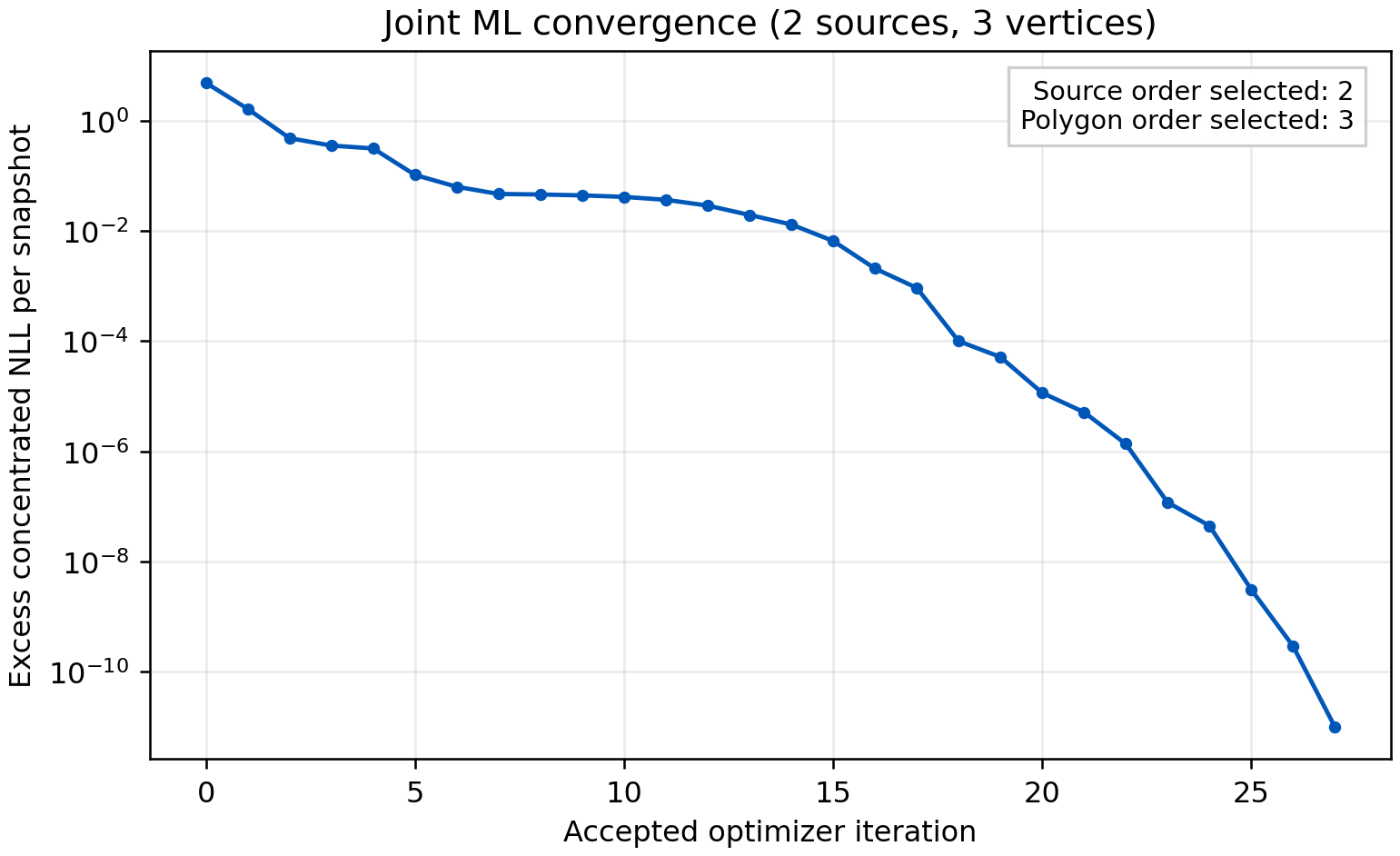}
\caption{\textcolor{black}{Decrease in negative log-likelihood for correct model order.}}
\label{Fig:poly}
\end{figure}

\begin{figure*}[!t]\centering
\includegraphics[width=.85\textwidth]{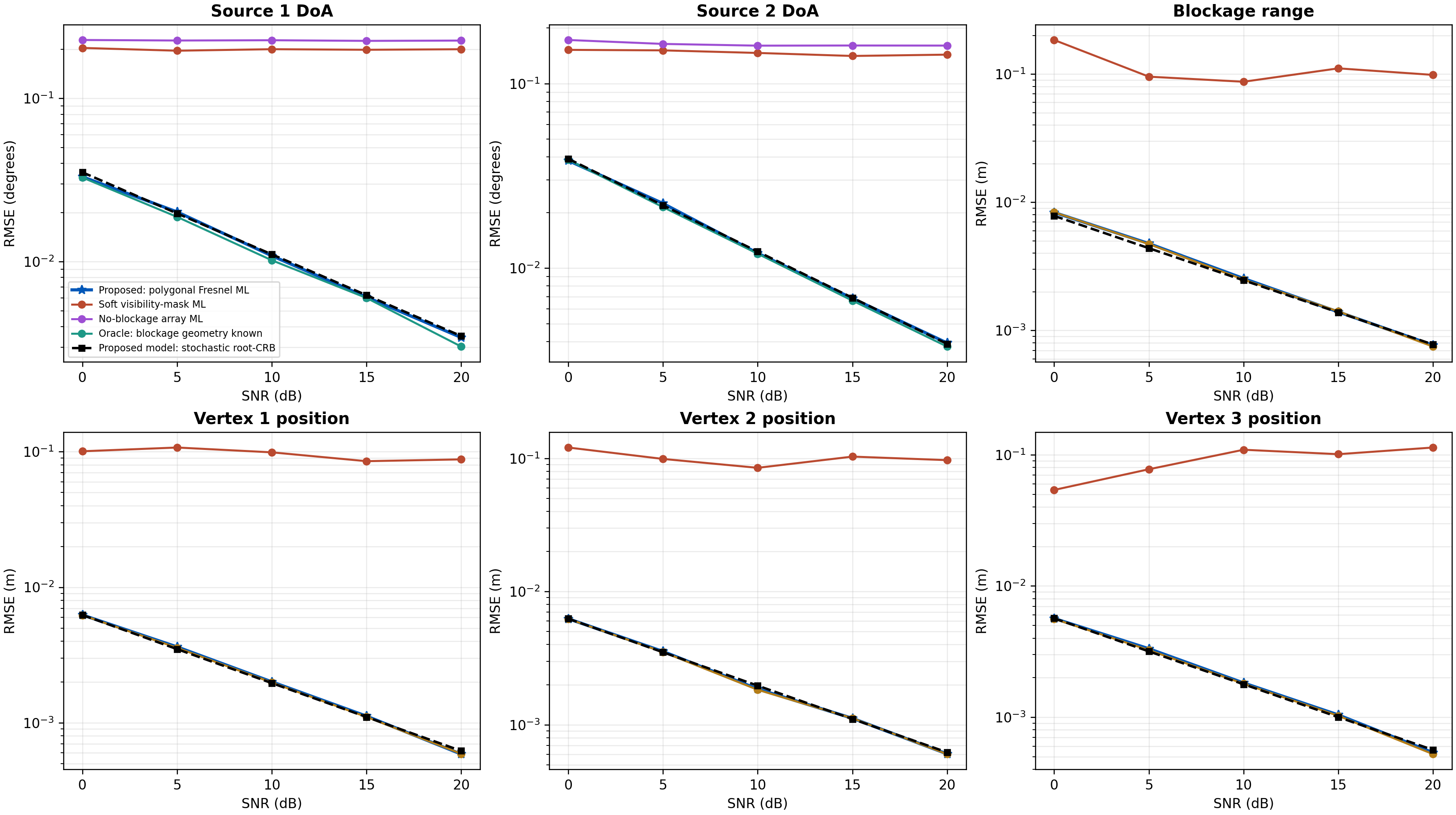}
\caption{\textcolor{black}{Joint polygonal RMSEs for source directions, range, and three vertex positions versus SNR, including proposed ML-based approach, stochastic root-CRB, mask and blockage-free baselines, and oracle performance. RMSE curves calculated for correct order selection. The two oracle references use either known blockage geometry (DoA performance curves) or known source DoAs (for range and vertex performance).}}
\label{Fig:shape_CRB}\end{figure*}

\textcolor{black}{We conduct 160 trials with 180
snapshots each at several SNRs, for two uncorrelated Gaussian sources of relative powers 1 and 0.8.
For a vertex $v$, the position RMSE is
$\sqrt{\mathbb E[(\widehat x_v-x_v)^2+
(\widehat y_v-y_v)^2]}$ and its root-CRB is
$\sqrt{[\operatorname{CRB}(\bm{\mu})]_{x_vx_v}+[
\operatorname{CRB}(\bm{\mu})]_{y_vy_v}}$.  
All observations are generated by the Babinet-normalized
coherent Fresnel response of the opaque triangle. The proposed method
uses a convex-polygon Fresnel integral evaluated by fan
triangulation and Gaussian quadrature, and estimates the two source
azimuths, blockage range, and six vertex coordinates after MDL model
order selection. For comparison, we implement the following: (i) a soft visibility-mask baseline that replaces
the coherent Fresnel-fringe factor by a smooth, real-valued
geometrical visibility/attenuation transition, but that estimates the same
source angles, range, and vertices with the same 
likelihood criterion; (ii) a method that ignores the presence of the blockage and estimates only the source angles assuming standard plane-wave
steering vectors (it cannot produce blockage shape or range estimates); and (iii) two oracle
references that use either the true object geometry when estimating
the source DoAs, or the true DoAs when estimating the
blockage range and vertices. Fig.~\ref{Fig:shape_CRB} shows the resulting RMSEs and associated CRBs. We see that the proposed approach yields performance essentially equivalent to the oracles, with a $6\times$ improvement over the baselines for the source DoA at 0~dB and two orders of magnitude better vertex-position estimates at 20~dB.}

\begin{figure}[!t]\centering
\includegraphics[width=.95\linewidth]{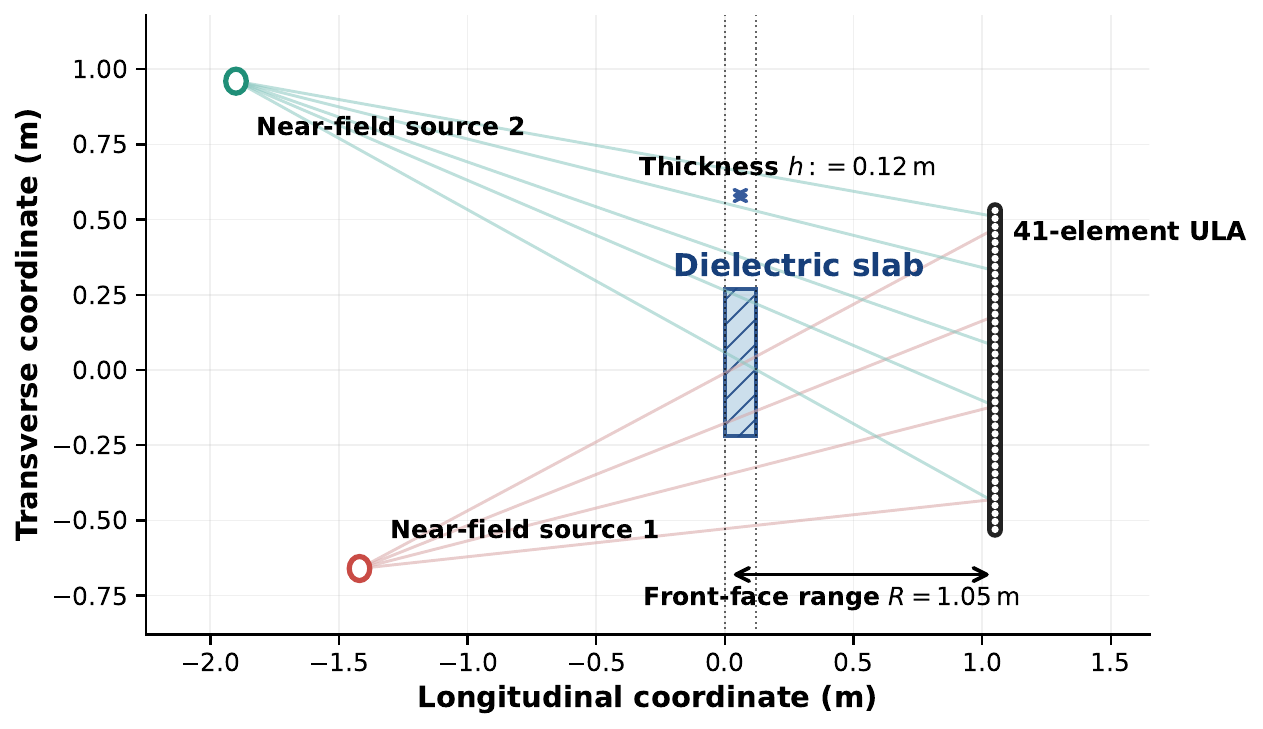}
%\vspace{-0.4cm}
\caption{\textcolor{black}{Geometry for near-field source localization and finite-thickness rectangular-blockage estimation. The slab has distinct source-facing range $R$ and axial thickness $h$.}}
%\vspace{-0.4cm}
\label{Fig:num}\end{figure}

\subsection{\textcolor{black}{Near-Field Sources and Blockage Depth Estimation}}\label{subsec:nf_rect_new}
\textcolor{black}{Finally, in Fig.~\ref{Fig:num} we consider two near-field sources with unknown DoAs $15^\circ$ and $-18^\circ$ and unknown ranges 2.55 and 3.10~m. A dielectric slab of width 0.48~m, transverse center 0.025~m, source-facing range $R=1.05$~m, thickness $h=0.12$~m, and refractive index $n_r=1.42+0.085j$ lies between the sources and a 41-element, $\lambda/2$-spaced ULA at 5.9~GHz. The unknown parameters are the source DoAs and ranges, and the slab range and thickness. Here $k_0\operatorname{Im}\{n_r\}h\approx1.26$, giving approximately $-11$~dB one-way field transmission, so the slab cannot be treated as opaque.}

\textcolor{black}{The Monte Carlo observations are generated with the local-wavenumber split-step model presented earlier, with six axial layers, nonparaxial angular-spectrum propagation, and a 512-point transverse FFT grid. The two source fields have cylindrical spreading proportional to $\exp(jk_0r)/\sqrt r$. For each SNR, independent observations consist of 80 snapshots of two uncorrelated circular complex-Gaussian source signals with relative powers 1 and 0.8, and the spatially white noise variance is chosen to achieve the desired SNR. A separate full-wave comparison below quantifies the effects of physical model mismatch.}

\begin{figure*}[!t]\centering
\includegraphics[width=\textwidth]{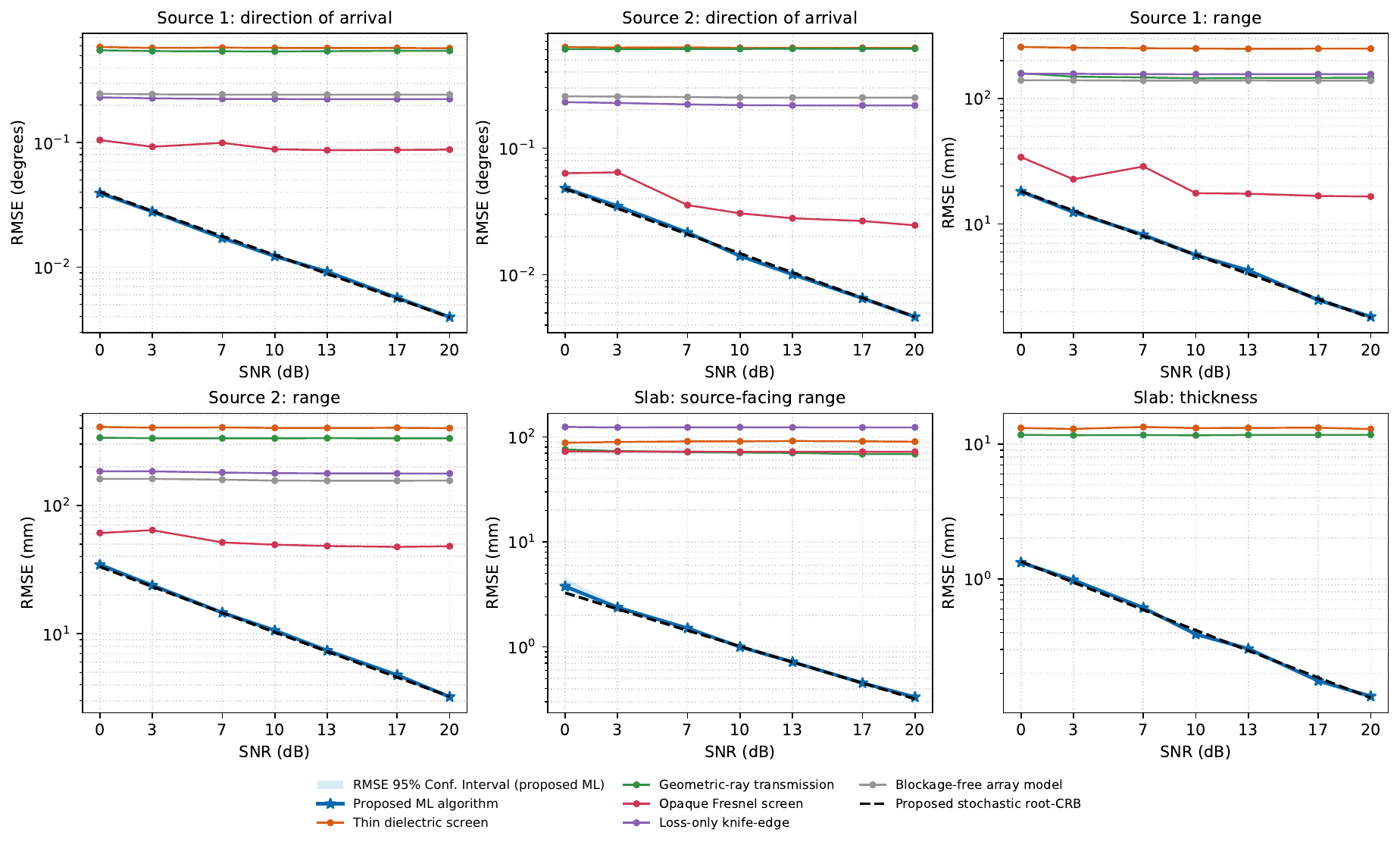}
\vspace{-0.4cm}
\caption{\textcolor{black}{Matched-model RMSE versus SNR for joint near-field source and finite-thickness slab estimation using the local-wavenumber split-step model. Each point uses 1000 independent trials and 80 snapshots. Shading indicates 95\% bootstrap uncertainty intervals for the proposed estimator's RMSE; the dashed black curves are nuisance-adjusted stochastic root-CRBs.}}
\label{ML_Fig}\end{figure*}

\begin{figure*}[!t]\centering
\includegraphics[width=.74\textwidth]{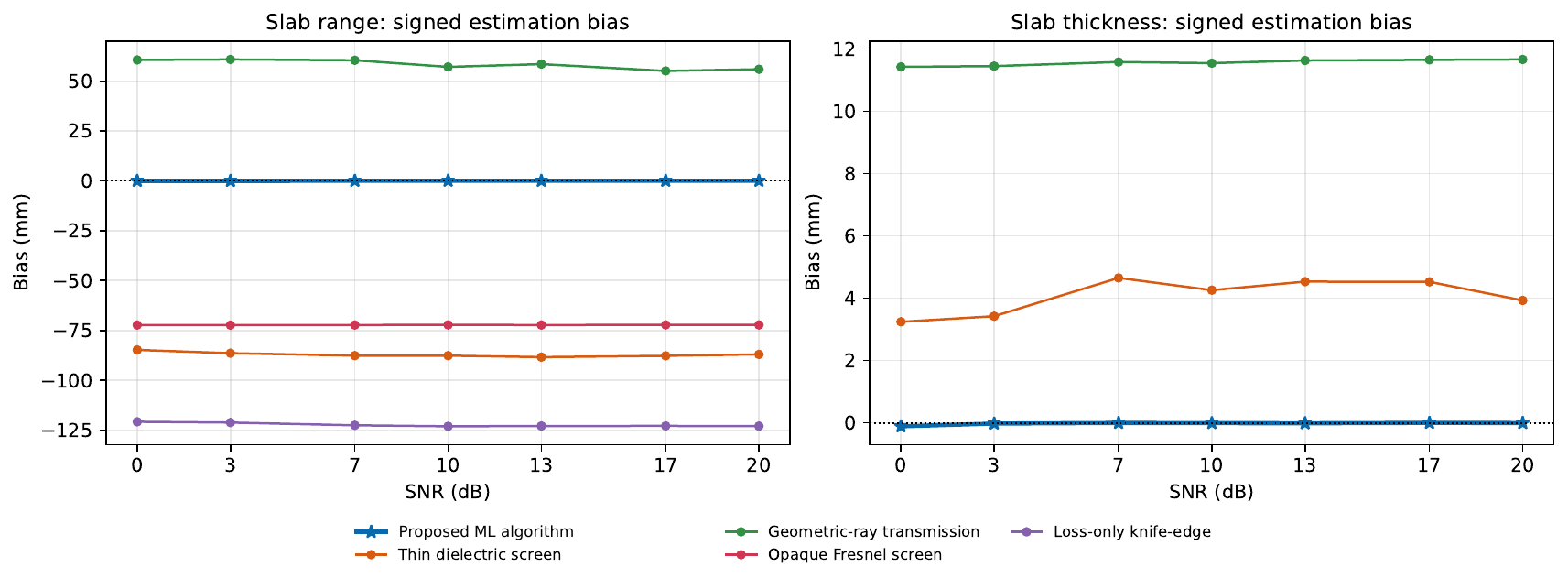}
\caption{\textcolor{black}{Matched-model slab-range and thickness bias for the proposed split-step estimator and alternative blockage models, with 1000 independent trials at each SNR. (These biases do not include the full-wave propagation mismatch quantified in Table~\ref{Tab:fullwave_modified}).}}
\label{CRB_Fig}\end{figure*}

\textcolor{black}{The proposed ML algorithm is 
compared to the following benchmarks: 
(i) a thin dielectric screen model that treats the object as
a single thin layer at its center. The layer introduces phase
shifts and attenuation according to
$\exp[jk_0(n_r-1)h]$, and Fresnel propagation is used to
calculate the received field;
(ii) a geometric-ray model that accounts for the distance
$\ell$ that each source-to-receiver ray travels through the
dielectric, using the transmission factor
$\exp[jk_0(n_r-1)\ell]$, but ignores diffraction around the
object's edges;
(iii) an opaque Fresnel screen that treats the object as
completely blocking the incident wave and calculates the
diffraction around its edges using Babinet's principle;
(iv) a knife-edge model that approximates the blockage loss
using the distance of the propagation path from the nearest
object edge, accounting only for attenuation, ignoring diffraction
phase and dielectric transmission; and
(v) a blockage-free model that assumes the conventional
near-field array response without accounting for the object. The knife-edge benchmark uses the standard approximation
for single-edge diffraction loss~\cite{ITURP526}: $L_{ke}(v)=0$ for $v\leq-0.78$ and otherwise}
\begin{equation}\label{eq:ked_loss_comparator_expanded}
\textcolor{black}{L_{ke}(v)=6.9+20\log_{10}
\!\left(\sqrt{(v-0.1)^2+1}+v-0.1\right)\ \mathrm{dB}.}
\end{equation}
\textcolor{black}{For source $k$ and sensor $m$, the loss-only benchmark uses
\begin{equation}
  \begin{aligned}
  v_{m,k}&=\sqrt{\frac{2}{\lambda R_{{\rm eff},k}}}
  \min\{\bar x_{m,k}-x_{\rm L},\ x_{\rm R}-\bar x_{m,k}\},\\
  R_{{\rm eff},k}&=\frac{L_k R}{L_k+R},
  \end{aligned}
  \label{eq:ked_v_explicit}
\end{equation}
where $x_{\rm L}$ and $x_{\rm R}$ are the left and right blockage edges and $\bar x_{m,k}$ is the transverse intersection of the unblocked source--receiver ray with the object plane (horizontal coordinate of $\mathbf c_{m,k}$ in~\eqref{eq:effective_fresnel_new}). Thus $v_{m,k}$ is positive for rays crossing the object interior and negative for rays passing outside an edge. The proposed ML algorithm and the
thin-screen and geometric-ray methods estimate all six 
unknowns. The opaque and loss-only models estimate the two source DoAs and ranges and the object range, but not the thickness.
The blockage-free method estimates the source DoAs and ranges only. \textcolor{black}{The variable $R$ refers to the array-to-source-facing-surface distance. In the thin-screen model, the equivalent transmission mask is placed at $h/2$ behind that surface, but $R$ is still the estimated front-surface range; thus the estimators use the same range reference when computing errors.}}

\textcolor{black}{Fig.~\ref{ML_Fig} shows the six-parameter RMSE comparison for the proposed model, using 1000 Monte Carlo trials and 80 snapshots at each SNR, and Fig.~\ref{CRB_Fig} provides a bias comparison. The proposed model achieves the stochastic root-CRB under its matched-model assumption, while the simplified alternatives exhibit model-dependent error floors. The full-wave results below provide a complementary check of the physical model mismatch.}

\begin{figure*}[!t]
\centering
\includegraphics[width=.75\textwidth]{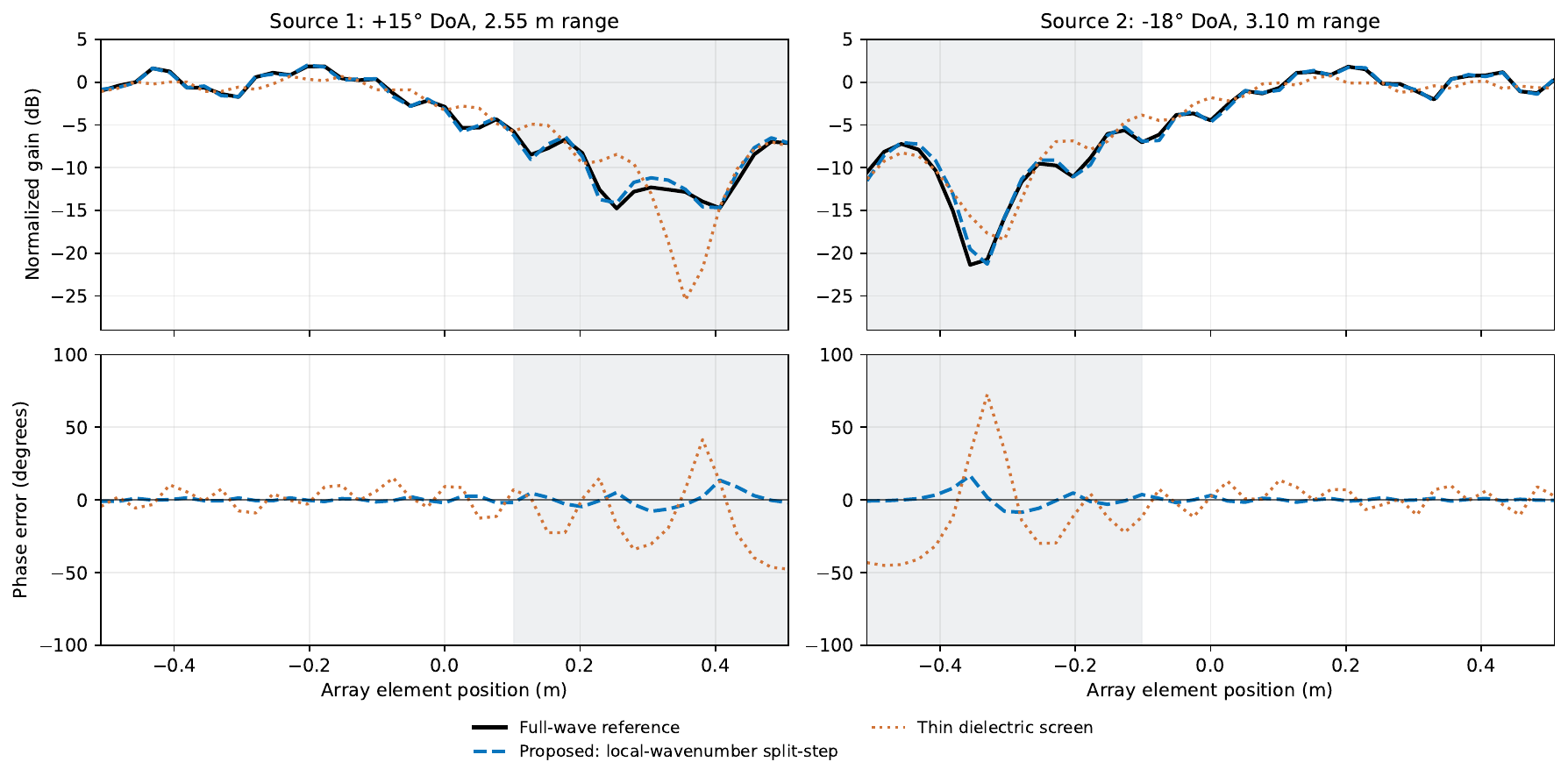}
\caption{\textcolor{black}{2D full-wave comparison of the out-of-plane electric field for the two near-field sources. Top: normalized complex-field gain $20\log_{10}|g_m|$ across array. Bottom: phase difference between approximate models and the full-wave field. Gray regions identify the shadowed array aperture.}}
\label{Fig:fullwave_modified}
\end{figure*}

\textcolor{black}{To assess the error due to propagation modeling rather than noise, we computed the array responses for this scenario using a 2D full-wave boundary-integral solver for a lossy dielectric rectangle. The solver models boundary reflections and corner diffraction and was checked against a Mie-series solution and an independent volume-integral calculation. We then compared the resulting complex responses for both out-of-plane field polarizations with the scalar models. With a refined 2048-point transverse grid, the proposed six-slice model gives a 3.6--6.0\% relative array-response error, compared with approximately 23--37\% for the simplified screen and ray models. Fig.~\ref{Fig:fullwave_modified} illustrates the excellent amplitude and phase agreement for the out-of-plane electric field.}

\textcolor{black}{As an additional check, we fit each approximate model to the noise-free full-wave fields using the same six source/slab parameters. Table~\ref{Tab:fullwave_modified} reports the largest absolute offsets over the two sources and polarizations. The proposed model has a maximum thickness error of 1.43~mm and a maximum slab-range error of 6.25~mm, compared with thickness errors of 6.6--12.8~mm and slab-range errors of 31.3--129~mm for the relevant benchmarks. These deterministic offsets provide a model-mismatch floor that is absent from the matched-model RMSE and bias curves of Figs.~\ref{ML_Fig} and~\ref{CRB_Fig}.}

\begin{table*}[!t]\centering
\caption{\textcolor{black}{Maximum relative field errors and absolute parameter-estimation
errors obtained by fitting the propagation models to noise-free
2D full-wave data, over both sources and both polarizations. Estimates were initialized at the true parameters, reported errors = resulting local minima.}}
\label{Tab:fullwave_modified}
\begin{tabular}{lccccc}\hline
Model & Field error & DoA (deg) & Source range (mm) & Slab range (mm) & Thickness (mm)\\\hline
ML split-step (proposed) & 6.0\% & 0.029 & 22.3 & 6.25 & 1.43\\
Thin dielectric screen & 23.8\% & 0.542 & 351 & 49.3 & 6.6\\
Geometric-ray model & 36.7\% & 0.622 & 324 & 31.3 & 12.8\\
Opaque Fresnel screen & 31.9\% & 0.055 & 102 & 74.8 & ---\\
Loss-only knife-edge & 26.7\% & 0.262 & 181 & 129 & ---\\\hline
\end{tabular}
\end{table*}

\subsection{Multiple Blockages at Different Ranges}
\begin{figure}[t!]
\centering
\includegraphics[width=.85\linewidth]{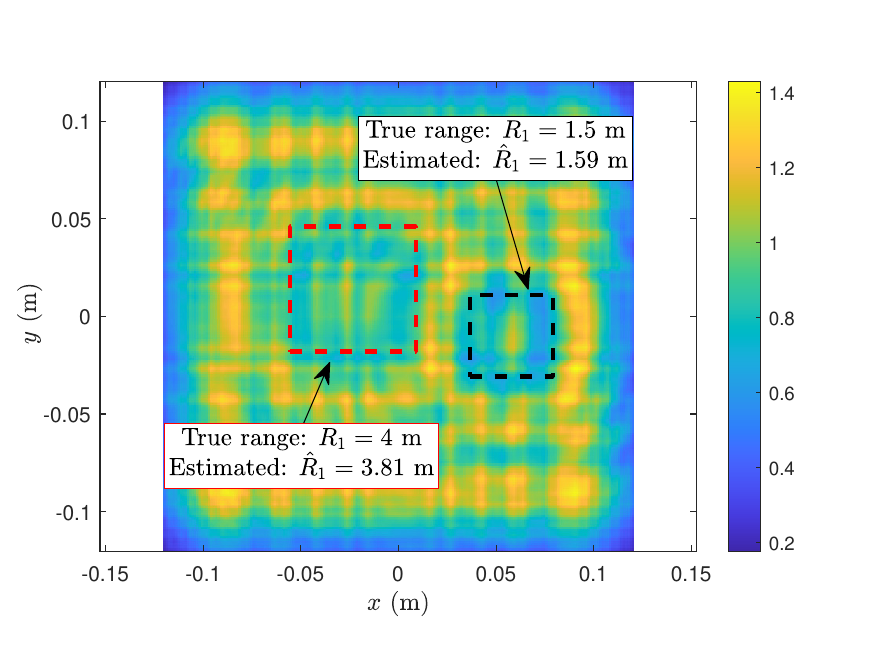}
\caption{\textcolor{black}{Receive-field magnitude for two laterally separated rectangular objects at different ranges. Range estimation assumes known shapes and an approximate additive response.}}
\label{Fig:ranges}
\end{figure}

Fig.~\ref{Fig:ranges} shows the magnitude of the complex field recorded
at a UPA when two laterally separated rectangular objects are present at
ranges $R_1 = 1.5$~m (black) and $R_2 = 4$~m (red). The received field
is generated with the full multi-blockage model described in
Section~II-F, with mutual shadowing and superposition. We estimate the ranges with an ML model that approximates the field as the sum
of the individual single-blockage responses and assumes the
rectangular shapes are known. Despite this simplification,
the ML algorithm returns accurate estimates of $\hat{R}_1 = 1.59$~m
and $\hat{R}_2 = 3.81$~m, demonstrating that the diffraction pattern encodes information
that can separate multiple laterally disjoint objects, and that
reasonable range estimates can be extracted from a single
diffraction snapshot, without echo timing or frequency
sweeps. \textcolor{black}{More complicated settings involving overlapping projections, unknown shapes, correlated diffraction, and closely spaced depths require sequential forward modeling and further analysis, which we defer to future work.}

\section{Conclusions}\label{sec:conclusion}
\textcolor{black}{We proposed a joint source-localization and environmental-parameter estimation framework based on a low-dimensional diffraction-aware array manifold. Its objective is not to replace full-wave EM modeling, but to retain the dominant geometry-dependent complex-field effects absent from conventional blockage-free or binary-visibility array responses. A common scalar formulation accommodates far- and near-field sources, opaque thin objects, and approximate finite-thickness dielectric/tilted extensions. Stochastic ML methods, CRBs, and MDL-type order selection allow the model to be used directly with array-processing tools. Numerical examples demonstrate locally near-statistically-efficient parameter estimation for polygonal and superelliptical shapes and finite-thickness rectangular objects. General sufficient identifiability conditions, consistent multi-object model-order selection, practical reduced-RF-chain architectures, robust global optimization, and broader full-wave and measured-data validation across different objects and materials remain important future directions.}

\bibliography{references}

\bibliographystyle{IEEEtran}

\end{document}